\documentclass[%
reprint,
superscriptaddress,
nofootinbib,
amsmath,amssymb,
aps,
prl,
floatfix,
twocolumn,
notitlepage
]{revtex4-2}
\usepackage{bbm}
\usepackage{lipsum}
\usepackage{graphicx}
\usepackage{dcolumn}
\usepackage{subfigure}
\usepackage{bm}
\usepackage{breakurl}
\usepackage{enumitem}
\usepackage{xcolor}
\usepackage[normalem]{ulem}
\usepackage{multirow}
\usepackage[multidot]{grffile}

\definecolor{myblue}{rgb}{0.153,0.322,0.706}
\usepackage[colorlinks,linkcolor=myblue,urlcolor=myblue,citecolor=myblue]{hyperref}
\usepackage{geometry}
\DeclareMathOperator*{\argmin}{arg\,min}
\newcommand{\abs}[1]{\left|#1\right|}
\def\multiset#1#2{\ensuremath{\left(\kern-.3em\left(\genfrac{}{}{0pt}{}{#1}{#2}\right)\kern-.3em\right)}}
\graphicspath{{./Pictures/13Dec2021}}

\usepackage[usenames,dvipsnames,table]{xcolor}
\usepackage{booktabs}

\begin{document}

\title{\large\bfseries Structural reducibility of hypergraphs}

\author{Alec Kirkley}
\email{alec.w.kirkley@gmail.com}
\affiliation{Institute of Data Science, University of Hong Kong, Hong Kong SAR, China}
\affiliation{Department of Urban Planning and Design, University of Hong Kong, Hong Kong SAR, China}
\affiliation{Urban Systems Institute, University of Hong Kong, Hong Kong SAR, China}

\author{Helcio Felippe}
\affiliation{Department of Network and Data Science, Central European University, 1100 Vienna, Austria}

\author{Federico Battiston}
\email{battistonf@ceu.edu}
\affiliation{Department of Network and Data Science, Central European University, 1100 Vienna, Austria}

\date{\today}

\begin{abstract}
Higher-order interactions provide a nuanced understanding of the
relational structure of complex systems beyond traditional pairwise
interactions. However, higher-order network analyses also incur more
cumbersome interpretations and greater computational demands than
their pairwise counterparts. Here we present an information-theoretic
framework for determining the extent to which a hypergraph
representation of a networked system is structurally redundant, and
for identifying its most critical higher orders of interaction that
allow us to remove these redundancies while preserving essential
higher-order structure. 
\end{abstract}

\maketitle

A wide variety of complex systems and relational data are
characterized by higher-order, non-dyadic
interactions~\cite{battiston2020networks,battiston2021physics,bianconi2021higher,bick2023higher,majhi2022dynamics}.
Such systems can be conveniently represented as hypergraphs,
collections of nodes representing fundamental units of a system that
are connected by hyperedges encoding interactions among an arbitrary
number of nodes~\cite{berge1984hypergraphs}. To investigate the
higher-order architecture of networked systems, new mathematical and
computational frameworks have been
proposed~\cite{benson2018simplicial, petri2018simplicial,
contisciani2022inference, di2024percolation}, revealing previously
unknown organizational principles and new emergent behaviours in
collective phenomena ranging from
contagions~\cite{iacopini2019simplicial, burgio2024triadic,
ferraz2024contagion} and diffusion~\cite{di2024dynamical} to
synchronization~\cite{skardal2019abrupt,millan2020explosive,zhang2023higher,anwar2024collective,majhi2024patterns}
and evolutionary dynamics~\cite{alvarez2021evolutionary,
civilini2024explosive,kumar2021evolution}.  Nevertheless, due to the
high dimensionality of many real-world hypergraphs, higher-order
network analyses are typically more computationally demanding and
complex than pairwise network analyses. Hence, it is important to
identify and exploit redundancies---which have been observed in
real-world
systems~\cite{lotito2022higher,larock2023encapsulation,landry2024simpliciality,gallo2024higher}---to
construct more compressed representations that retain the key
structural heterogeneity present in a system's original higher-order
structure. 

Inspired by related work in the context of multilayer
networks~\cite{de2015structural,de2016spectral,santoro2020algorithmic},
here we provide a simple and principled information theoretic solution
to identify the \emph{structural reducibility} of a hypergraph---the
extent to which a hypergraph provides redundant information about a
system's relational structure---and remove these redundancies to
create a reduced representation that retains its critical higher-order
structure. Our method is interpretable, computationally efficient, and
can be generalized to capture the reducibility of hypergraphs when
viewed at different scales. We test our framework on a variety of
synthetic network models, showcasing its wide applicability and
robustness to different sources of statistical noise.  Finally, we
apply the framework to a corpus of real-world higher-order systems
from various application domains, finding that many of these systems
can be substantially structurally reduced. 

\textit{Hypergraph reducibility}---Let $G=\{G^{(\ell)}\}_{\ell\in
\mathcal{L}}$ be a hypergraph with $L$ unique (but not necessarily
consecutive) layers~$G^{(\ell)}$ indexed by $\ell$, each
layer~$G^{(\ell)}$ containing all hyperedges of size~$\ell$ from $G$.
Let \mbox{$\mathcal{L}=\{\ell_1,\dots,\ell_L\}$} denote the set of $L$
unique layer indices. For example, a hypergraph~$G$ with only
layers~$\ell=2$ and $\ell=5$ would have $G=\{G^{(2)},G^{(5)}\}$ and
$\mathcal{L}=\{2,5\}$. We consider $G^{(\ell)}$ as a set of
undirected, sorted tuples of size~$\ell$ with no repeated entries, and
let $E^{(\ell)}=\abs{G^{(\ell)}}$ be the number of hyperedges in
$G^{(\ell)}$ (in other words, $G$ is a simple, undirected hypergraph).
There are ${N\choose \ell}$ possible undirected, sorted tuples of size
$l$ so \mbox{$E^{(\ell)}\leq {N\choose \ell}$}. We also let $G^{(k\to
\ell)}$ be the projection of layer~\mbox{$k>\ell$} onto order~$\ell$,
which extracts all unique $\ell$-tuples nested within the $k$-tuples
in $G^{(k)}$. For example, if $G^{(3)}=\{(0,1,2),(0,2,4)\}$, we would
have $G^{(3\to 2)}=\{(0,1),(0,2),(1,2),(0,4),(2,4)\}$. We define
\mbox{$E^{(k\to \ell)}=\abs{G^{(k\to \ell)}}$}, similarly to the
unprojected layers, and use the convention $G^{(\ell\to
\ell)}=G^{(\ell)}$. 

The structural reducibility of a hypergraph $G$ can be defined based
on the overlap among its pairs of
layers~$(G^{(k)},G^{(\ell)}),~k,\ell\in\mathcal{L}$, where overlap is
defined based on the projection of each layer to the lower order of
the two. This convention is required because higher order hyperedges
have unique projections onto lower order interactions (as defined
above), but one cannot conversely determine higher order structure
uniquely from lower order structure
alone~\cite{battiston2020networks}. Higher overlap among the layers
indicates higher structural redundancy among different orders of
hyperedges, suggesting a higher structural reducibility for the
hypergraph $G$. Formally, we can define the overlap of the layers
indexed by $k,\ell$ as $E^{(k \cap \ell)}=\abs{G^{(k\to
\text{min}(k,\ell))}\cap G^{(\ell\to \text{min}(k,\ell))}}$, which is
the number of hyperedges the two layers share when they are projected
to the lower order of the two layers. 

Our proposed reducibility measure reflects the extent to which a
hypergraph $G$ can be \emph{compressed} in an information theoretic
sense when we exploit the structural redundancy among its hyperedge
layers of different orders (i.e. their layer overlaps). To formalize
this concept mathematically, we consider transmitting the hypergraph
$G$ to a receiver using two different schemes. In the first (na\"ive)
scheme we transmit each of $G$'s layers $G^{(\ell)}$ individually. In
the second scheme, we first transmit a set of $R\leq L$
``representative'' layers $\mathcal{R}\subseteq \mathcal{L}$ which
capture most of the heterogeneity in the hyperedge structure in $G$,
then we transmit each remaining layer $\ell\in \mathcal{L}\setminus
\mathcal{R}$ as a noisy copy of a representative layer $r(\ell)\in
\mathcal{R}$. A similar concept has been employed to compress
multilayer network structures~\cite{kirkley2023compressing} and sets
of network partitions~\cite{kirkley2022representative} using
intermediate representative structures. 

\begin{figure}[t]
\includegraphics[width=0.50\textwidth]{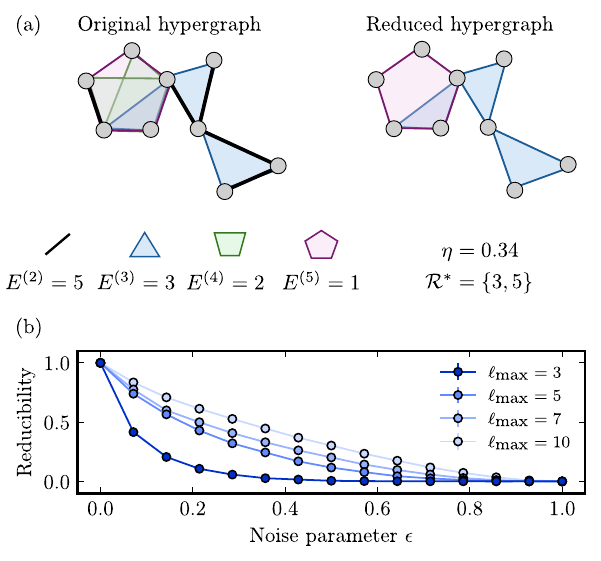}
\caption{
    Structural reducibility.
    (a)~Hypergraph containing layers $\mathcal{L}=\{2,3,4,5\}$ of size
    $E^{(\ell)}$, which is reducible to an optimal representative
    layer set $\mathcal{R}^*=\{3,5\}$ with reducibility $\eta=0.34$
    (Eq.~(\ref{eq:reducibility})).
    (b)~Reducibility of a noisy nested hypergraph, with noise
    parameter $\epsilon$ determining the fraction of randomized
    hyperedges, for various hypergraph
    dimensions~$\ell_{\textrm{max}}$. 
}
\label{fig:fig1_diagram_eta}
\end{figure}

We assume that the receiver knows the orders~$\mathcal{L}$ of the
layers and the number of hyperedges~$E^{(\ell)}$ for each $\ell\in
\mathcal{L}$---specifying these counts incurs a comparatively
negligible information cost anyway. Since each layer $G^{(\ell)}$ has
${{N\choose \ell}\choose E^{(\ell)}}$ possible configurations of its
hyperedges when $E^{(\ell)}$ is known, we need to send a bitstring of
length approximately equal to $\log_2 {{N\choose \ell}\choose
E^{(\ell)}}$ bits to fully specify which configuration corresponds to
$G^{(\ell)}$. The na\"ive transmission of $G$ as individual layers
therefore requires an information content of 
\begin{align} 
    H_0 = \sum_{\ell\in \mathcal{L}}\log {{N\choose \ell}\choose E^{(\ell)}}
\end{align} 
bits, using the convention $\log\equiv \log_2$ for brevity. 

A better way to transmit $G$ is to exploit the overlaps among layers
of hyperedges of different sizes to save information. To do this, we
first transmit a representative subset of $R\leq L$ layers indexed by
$\mathcal{R}\subseteq \mathcal{L}$, which incurs a cost of $\sum_{r\in
\mathcal{R}}\log {{N\choose r}\choose E^{(r)}}$ bits. Then we transmit
each remaining layer~$\ell\in\mathcal{L}\setminus\mathcal{R}$ by:
(a)~transmitting a representative layer~$r(\ell)\in \mathcal{R}$ of a
higher order, costing us $\log R$ bits as there are $R$ layers to
choose from; (b)~transmitting the overlap~$E^{(r(\ell)\cap \ell)}$
among the layer $\ell$ and its representative $r(\ell)$, costing us
$\log (E^{(r(\ell)\to \ell)}+1)$ bits as $E^{(r(\ell)\cap \ell)}\in
[0,E^{(r(\ell)\to \ell)}]$; and (c)~transmitting layer~$\ell$ given
the constraints imposed by the overlap value $E^{(r(\ell)\cap \ell)}$.
If $E^{(r(\ell)\cap \ell)}$ of the possible $E^{(r(\ell)\to \ell)}$
edges in $G^{(r(\ell)\to \ell)}$ are present in $G^{(\ell)}$, and
$E^{(\ell)}-E^{(r(\ell)\cap \ell)}$ of the ${N\choose
\ell}-E^{(r(\ell)\to \ell)}$ possible edges absent from
$G^{(r(\ell)\to \ell)}$ are present in $G^{(\ell)}$, then we require
$\log {E^{(r(\ell)\to \ell)}\choose E^{(r(\ell)\cap \ell)}}{{N\choose
\ell}-E^{(r(\ell)\to \ell)}\choose E^{(\ell)}-E^{(r(\ell)\cap \ell)}}$
bits to specify $G^{(\ell)}$ given knowledge of $G^{(r(\ell))}$ and
$E^{(r(\ell) \cap \ell)}$. Steps~(a) and (b) incur negligible
information costs compared to step~(c) and can be ignored. The total
information content of this scheme is then 
\begin{align}\label{eq:HGR}
H_G(\mathcal{R}) &= \sum_{r\in \mathcal{R}}\log {{N\choose r}\choose E^{(r)}} \\
&+ \sum_{\ell\in \mathcal{L}\setminus \mathcal{R}}\log {E^{(r(\ell)\to \ell)}\choose E^{(r(\ell)\cap \ell)}}{{N\choose \ell}-E^{(r(\ell)\to \ell)}\choose E^{(\ell)}-E^{(r(\ell)\cap \ell)}}\nonumber
\end{align}
bits. For a minimal information cost, the representative
layer~$r(\ell)\in\mathcal{R}$ for each layer~$\ell\notin\mathcal{R}$
can be assigned as
\begin{align}
r(\ell) = \argmin_{r\in \mathcal{R},r>\ell} \left\{ \log {E^{(r\to \ell)}\choose E^{(r\cap \ell)}}{{N\choose \ell}-E^{(r\to \ell)}\choose E^{(\ell)}-E^{(r\cap \ell)}} \right\}.
\end{align}

The information cost~$H_G(\mathcal{R})$ of this transmission scheme
depends on which layers~$\mathcal{R}\subseteq \mathcal{L}$ are
selected as representatives---the better the layers $\mathcal{R}$
capture the heterogeneity in the hyperedge structure in $G$, the lower
the information cost $H_G(\mathcal{R})$. Thus, to maximize compression
from layer overlap, we must find the optimal set of representative
layers $\mathcal{R}^\ast$ according to
\begin{align}\label{eq:Rstar}
\mathcal{R}^{\ast} = \argmin_{\mathcal{R}\subseteq \mathcal{L}} \left\{H_G(\mathcal{R})\right\}. 
\end{align}
We will describe shortly how to solve this optimization problem.

The optimal information~cost~$H_G(\mathcal{R}^\ast)$ of this
transmission scheme is bounded in the interval
$H_{\ell_{\text{max}}}\leq H_G(\mathcal{R}^\ast)\leq H_0$, where
\begin{align}
H_{\ell_{\text{max}}}=\log{{N\choose \ell_{\text{max}}}\choose E^{(\ell_{\text{max}})}}.    
\end{align}
The lower bound follows from always minimally needing to transmit the
top layer of $G$ as a representative layer, at a cost
$H_{\ell_{\text{max}}}$, and the upper bound follows from
$H_0=H_G(\mathcal{R}=\mathcal{L})$ being in the solution space over
which we minimize $H_G$ to find $\mathcal{R^\ast}$. Therefore,
exploiting layer overlap always provides compression relative to the
na\"ive transmission of layers independently. The reducibility of the
hypergraph~$G$ can then be computed based on the extent to which $G$
can be compressed relative to the baseline cost of $H_0$ bits. 

Using these bounds we can construct a properly normalized
\emph{structural reducibility} measure~$\eta$ for a hypergraph $G$ as
\begin{align}\label{eq:reducibility}
\eta = \frac{H_0-H_G(\mathcal{R}^\ast)}{H_0-H_{\ell_{\text{max}}}},    
\end{align}
which satisfies $\eta\in [0,1]$. If $G$ is maximally
compressible---i.e., is a nested hypergraph where all layers
$G^{(\ell)}$ with $\ell<\ell_{\text{max}}$ are given by
$G^{(\ell_{\text{max}}\to \ell)}$---then we have
$\mathcal{R}^\ast=\{\ell_{\text{max}}\}$ and
$H_G(\mathcal{R}^\ast)=H_{\ell_{\text{max}}}$, thus $\eta=1$. On the
other hand, we have an information cost $H_G(\mathcal{R}^\ast) \approx
H_0$ when $G$ is highly incompressible (i.e. has little to no
structural overlap among its layers), as there is little shared
information that can be exploited to improve on the na\"ive
information cost of $H_0$, thus $\eta\approx 0$. 

In the Supplemental Material~\cite{prlSM}, we discuss extending our reducibility
concept to understand the structural redundancy of multiscale
coarse-grainings of hypergraphs (Sec.~\hyperlink{supp1}{I}), as well
as individual hypergraph layers (Sec.~\hyperlink{supp5}{V}) and
individual hyperedges (Sec.~\hyperlink{supp6}{VI}).

By solving Eq.~(\ref{eq:Rstar}) to maximize compression of $G$, we can
also obtain a compressed hypergraph representation for $G$ given by
$G_{\text{red}}=\{G^{(r)}\}_{r\in \mathcal{R}^\ast}$, which captures
the critical higher-order structure of $G$ while removing structurally
redundant layers. In Fig.~\ref{fig:fig1_diagram_eta}(a) we show an
example hypergraph with layers $\mathcal{L}=\{2,3,4,5\}$ and its
corresponding reduced representation of layers
$\mathcal{R}^\ast=\{3,5\}$, giving a reducibility value of
$\eta=0.34$.  

\textit{Optimizing reducibility}---To identify the optimal
representative layers $\mathcal{R}^{\ast}$, when $L\lesssim 30$---a
value satisfied by most real hypergraph
datasets~\cite{lucas2024functional}---we can use a simple brute force
search. Letting $\ell_{\text{max}}=\text{max}(\mathcal{L})$ be the
largest hyperedge size in $G$, we must have $\ell_{\text{max}}\in
\mathcal{R}$, since there is no higher layer from which to transmit
$G^{(\ell_{\text{max}})}$. Therefore, we have $2^{L-1}$ possible
subsets of representatives $\mathcal{R}$ that we must search through.
In our method we must first compute $E^{(k\to \ell)}$, $E^{(k\cap
\ell)}$, and $E^{(\ell)}$ for all pairs of layers~$k,\ell$ with
$k>\ell$. For layer pairs with small values $k,\ell \lesssim 10$, we
can directly compute $E^{(k\to\ell)}$ and $E^{(k\cap \ell)}$ using the
projection $G^{(k\to \ell)}$ and the lower layer $G^{(\ell)}$. For
larger $k,\ell$ values, however, we cannot compute the
projection~$G^{(k\to \ell)}$ directly. In this case, we can compute
$E^{(k\cap \ell)}$ by iterating through the edges~\mbox{$e_t\in
G^{(k)}$} in a fixed order, for each edge~$e_t$ checking its overlaps
\mbox{$o(e_t)=\{e_t\cap e_\tau:\tau < t\}$} with all previously
checked edges. Then, we can compute the number of new projected tuples
that $e_t$ contributes to $E^{(k\to \ell)}$ as ${k\choose
\ell}-E^{(o(e_t)\to \ell)}$, where $E^{(o(e_t)\to \ell)}$ is the
number of unique subtuples of size~$\ell$ within the set of
overlapping tuples~$o(e_t)$, which can be computed recursively using
the same approach. When computing the projection $G^{(k\to l)}$ is
infeasible, we also must compute $E^{(k\cap\ell)}$ in a more efficient
way. We do this by iterating over hyperedges~$e_k\in G^{(k)}$ and
incrementing $E^{(k\cap \ell)}$ for each edge $e_\ell\in G^{(\ell)}$
that fully overlaps with $e_k$, removing $e_l$ from $G^{(\ell)}$
afterwards. 

We can then compute a matrix $M$ such that
\begin{align}
M[k,\ell] = \log {E^{(k\to \ell)}\choose E^{(k\cap \ell)}}{{N\choose \ell}-E^{(k\to \ell)}\choose E^{(\ell)}-E^{(k\cap \ell)}}   
\end{align}
for layer pairs~$(k,\ell),~k>\ell$. The computation of certain
conditional entropies $M[k,\ell]$ is often the computational
bottleneck in practice, and takes roughly $O((E^{(k)})^2)$ runtime
for large $k,\ell$ where projection is intractable and
$O({k\choose \ell}E^{(k)})$ runtime when using direct
projection. We also compute a vector $Q$ with entries
\begin{align}
Q[\ell] = \log {{N\choose \ell}\choose E^{(\ell)}}    
\end{align}
storing the individual layer information costs for all layers~$\ell$.
Then, for each valid subset~$\mathcal{R}\subseteq \mathcal{L}$, we can
compute
\begin{align}
H_G(\mathcal{R}) = \sum_{r\in\mathcal{R}}Q[r] +\sum_{\ell\in \mathcal{L}\setminus\mathcal{R}}M[r(\ell),\ell],   
\end{align}
where \mbox{$r(\ell) = \argmin_{r\in
\mathcal{R}}\left\{M[r,\ell]\right\}$}. We then select
$\mathcal{R}^\ast$ as the representative subset that achieves the
minimum value of $H_G(\mathcal{R})$. There are $2^{L-1}$
subsets~$\mathcal{R}$ to check, and each takes in the worst case
$O(L)$ operations to compute $r(\ell)$ for each layer~$\ell\in
\mathcal{L}\setminus \mathcal{R}$. The iteration over $\mathcal{R}$
thus has a time complexity of roughly $O(L^22^{L-1})$, which in
practice is tractable for most real-world datasets with $L\lesssim 30$
(see Table~\ref{tab:datasets}). 

\begin{figure}[t]
\includegraphics[width=0.50\textwidth]{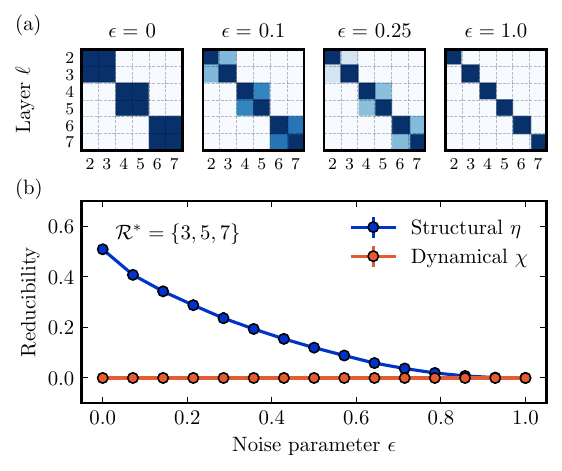}
\caption{
    Comparison of structural and dynamical~\cite{lucas2024functional}
    reducibility measures. 
    (a)~Pairwise layer similarity matrices of block-nested hypergraphs
    at increasing levels of noise $\epsilon$.  Hypergraph
    layers~$\ell=2$, $4$, and $6$ are fully nested within $\ell=3$,
    $5$, and $7$, respectively, and the similarity is lost as
    $\epsilon$ increases.
    (b)~Structural and dynamical reducibility measures against all
    $\epsilon$ values.  The dynamical reducibility does not detect any
    compressibility between layers. The structural reducibility
    uncovers both the structural redundancies and the planted, optimal
    representative layers~$\mathcal{R}^*=\{3,5,7\}$.     
}
\label{fig:fig2_structural-functional}
\end{figure}

For systems with many layers $\mathcal{L}$, optimizing over subsets
$\mathcal{R}\subseteq \mathcal{L}$ through direct enumeration is
unfeasible. In such cases, we use an approximate greedy method for
identifying $\mathcal{R}^\ast$ to compute the structural reducibility.
Starting with $\mathcal{R}=\{\ell_{\text{max}}\}$, we can iteratively
add the best layer~$\ell$ to $\mathcal{R}$ following the rule
\mbox{$\ell=\argmin_{\ell\in
\mathcal{L}\setminus\mathcal{R}}\{H_G(\{\ell\}\cup \mathcal{R})\}$}
until $\mathcal{R}=\mathcal{L}$. Then, we choose among the explored
solution candidates to find the representative layer set giving the
lowest information cost $H_G$. In practice, this approximation always
obtains accurate results (see Sec.~\hyperlink{supp4}{IV} of the
Supplemental Material~\cite{prlSM}), significantly speeding up the computation and
making our method available to otherwise intractable datasets.

\textit{Reducibility of synthetic hypergraphs}--- To validate our
approach, we investigate our method on synthetic hypergraphs with
tunable structure. First, we consider nested hypergraphs, where all
lower-order interactions are fully incapsulated into those of higher
order.  Noisy nested hypergraphs are nested hypergraph where a noise
parameter $\epsilon$ determines the fraction of its hyperedges to be
rewired, replacing each selected hyperedge with a hyperedge of the
same order drawn uniformly at random. In
Fig.~\ref{fig:fig1_diagram_eta}(b) we plot the results of this test,
averaged over ten realizations of the randomness for each value of
$\epsilon$. Standard errors (vanishingly small) are shown as error
bars. We can observe that $\eta=1$ indicates complete reducibility
when $\epsilon=0$ for each fully nested hypergraph, and that $\eta$
decreases smoothly as the hypergraph becomes noisier, eventually
bottoming out at $\eta=0$ for purely random hypergraphs
($\epsilon=1$). As the system gets larger ($\ell_{\rm max}$ increases)
we observe greater reducibility values, since a larger fraction of the
layers are structurally redundant due to being nested within the top
layer~$\ell_{\rm max}$.

In a follow-up experiment we examine the reducibility of more general
synthetic hypergraphs with nested structure. We start by generating
three planted representative layers $G^{(3)}$, $G^{(5)}$, and
$G^{(7)}$ on $N=100$~nodes with \mbox{$E^{(3)}$, $E^{(5)}$,
$E^{(7)}=1740$, $1050$, and $50$} hyperedges respectively, drawn
uniformly at random without replacement. We then generate the
layers~$G^{(2)}$, $G^{(4)}$, and $G^{(6)}$ as noisy versions of the
projected layers~$G^{(3\to 2)}$, $G^{(5\to 4)}$, and $G^{(7\to 6)}$
respectively by selecting a fraction~$\epsilon$ of the hyperedges in
each projected layer randomly and replacing these hyperedges with
those of the same size drawn uniformly at random. In
Fig.~\ref{fig:fig2_structural-functional} we plot layer-layer
similarity matrices showing the network normalized mutual
information~\cite{felippe2024network} (over arbitrary tuple sizes) for
pairs of layers in the generated hypergraphs, illustrating the effect
of $\epsilon$ on the nested structure. As $\epsilon$ increases we see
a smooth decrease in the structural reducibility~$\eta$, with our
method able to infer the planted set of representative
layers~$\mathcal{R}^\ast$.  We compare our results with dynamical
reducibility~\cite{lucas2024functional}, which reduces structure based
not on topological overlap but on the collective behavior supported by
the hypergraph, which remains close to zero and does not detect any
change in this simple but nuanced hypergraph structure. All results
are averaged over 20 realizations of the randomness at each
$\epsilon$. Section~\hyperlink{supp2}{II} of the Supplemental
Material~/cite{prlSM}
further investigates the reducibility of hypergraphs with tunable
nested interactions.

\begin{table}[t!]
    \centering
    \scriptsize
    \rowcolors{2}{white}{gray!10}
    \begin{tabular}{lrrrcc} 
    \toprule 
    \textbf{Dataset}         & \multicolumn{1}{c}{$N$}  & \multicolumn{1}{c}{$E$} &  $\ell_{\rm max}$ &    $\mathcal{R^*}$ & $\eta$ \\
    \midrule 
    \fontsize{6}{7}\selectfont{coauth-mag-geology\_1980} & 1674 &  903 &         18 &        \{3, 5, 9, 18\}  & 0.03 \\ 
    \fontsize{6}{7}\selectfont{coauth-mag-geology\_1981} & 1075 &  547 &         29 &         \{4, 5, 8, 29\} & 0.01 \\
    \fontsize{6}{7}\selectfont{coauth-mag-geology\_1982} & 1878 &  987 &         26 &         \{4, 6, 26\}   & 0.02 \\ 
    \fontsize{6}{7}\selectfont{coauth-mag-geology\_1983} & 1734 &  883 &         36 &         \{4, 36\}   & 0.03$^{\dagger}$ \\
    \fontsize{6}{7}\selectfont{kaggle-whats-cooking}     & 6714 & 39224 &         65 &  {\{6,8,9,11,65\}} & 0.04$^{\dagger}$ \\ 
    \fontsize{6}{7}\selectfont{contact-high-school}      & 327  &  7818 &          5 &              \{3, 5\} & 0.13 \\ 
    \fontsize{6}{7}\selectfont{contact-primary-school}   & 242  & 12704 &          5 &              \{4, 5\} & 0.09 \\ 
    \fontsize{6}{7}\selectfont{hospital-lyon}            & 75   &  1824 &          5 &              \{4, 5\} & 0.11 \\ 
    \fontsize{6}{7}\selectfont{hypertext-conference}     & 113  &  2434 &          6 &            \{3, 5, 6\} & 0.06 \\ 
    \fontsize{6}{7}\selectfont{invs13}                   & 92   &  787 &          4 &              \{3, 4\} & 0.05 \\
    \fontsize{6}{7}\selectfont{invs15}                   & 217  &  4909 &          4 &              \{3, 4\} & 0.10 \\ 
    \fontsize{6}{7}\selectfont{science-gallery}          & 410  &  3350 &          5 &              \{3, 5\} & 0.16 \\ 
    \fontsize{6}{7}\selectfont{sfhh-conference}          & 403  &  10541 &          9 &              \{4,9\} & 0.10 \\ 
    \fontsize{6}{7}\selectfont{malawi-village}           & 84   &  431 &          4 &              \{3, 4\} & 0.19 \\ 
    \fontsize{6}{7}\selectfont{dawn}                     & 2290 & 138742 &         16 &        \{6,7,13,16\} & 0.15 \\ 
    \fontsize{6}{7}\selectfont{ndc-classes}              & 628  &  796 &         39 &  \parbox{2.0cm}{\{4, 9, 10, 13, 14,23,27,39\}} & 0.40$^{\dagger}$ \\ 
    \fontsize{6}{7}\selectfont{ndc-substances}           & 3414 &  6471 &        187 &       \{5, 40, 187\}                              & 0.31$^{\dagger}$ \\
    \fontsize{6}{7}\selectfont{email-enron}              & 143  &  1459 &         37 &  {\{4,6,11,12,37\}} & 0.17$^{\dagger}$ \\ 
    \fontsize{6}{7}\selectfont{email-eu}                 & 986  &  24520 &         40 &  \parbox{2.0cm}{\{5,7,8,9,10,11, 12,13,27,39,40\}}  & 0.19$^{\dagger}$ \\ 
    \fontsize{6}{7}\selectfont{tags-ask-ubuntu}          & 3021 &  145053 &          5 &                \{5\} &    0.17 \\ 
    \fontsize{6}{7}\selectfont{tags-math-sx}             & 1627 &  169259 &          5 &                \{5\} &    0.26 \\
    \bottomrule
    \end{tabular}
    \caption{Structural reducibility of empirical datasets. Daggers
    denote the usage of greedy minimization for obtaining $\eta$ and
    $\mathcal{R}^*$.  The greedy scheme produced identical results to
    the exact scheme for all networks with \mbox{$\ell_{\rm max}\lesssim 30$}.
    }
    \label{tab:datasets}
\end{table}

We also examine our proposed multiscale reducibility measure in a
similar experimental setting. For an~$N=10^4$~node system, we
synthesize each random hypergraph using a planted community
partition~$\bm{b}$ by generating each of the $E^{(\ell)}$ hyperedges
in layer~$\ell=2,\dots,10$ as follows: (1)~choose a random node~$i$ to
start a hyperedge~$e$; (2)~add $\ell-1$ nodes to $e$, drawing each
node (without replacement) from the same community as $i$ with
probability~$1-p$ and from a different community with probability~$p$.
The result is a hypergraph that is more tightly clustered under the
planted node partition~$\bm{b}$ as $p\to 0$ and uncorrelated with
$\bm{b}$ as $p\to 1$. All results are averaged over ten realizations
of the randomness at each $p$, and separate experiments are run for
\mbox{$B=\{50,200,1000\}$}~equally-sized communities in $\bm{b}$. We
observe that the standard reducibility of Eq.~(\ref{eq:reducibility})
cannot detect any changes in the mesoscale nested structure, while the
multiscale reducibility of Eq.~(S3) 
in the Supplemental Material~\cite{prlSM}
exhibits a
smooth descent as $p$ increases. For larger $B$, we see that the
communities in $\bm{b}$ become smaller and the hypergraph becomes less
reducible under the coarse-graining~$\bm{b}$, approaching the standard
structural reducibility value.  

\textit{Reducibility of real networks}---Finally, we apply our
reducibilty method to a range of real higher-order
networks~\cite{Landry_XGI_A_Python_2023}, as shown in
Table~\ref{tab:datasets}. We find a great variety in the reducibility
of these systems, with many systems most parsimoniously represented by
only a small subset $\mathcal{R}^\ast$ of their layers.  In the
Supplemental Material~\cite{prlSM}, we further investigate the nested organization
of real-world systems in Sec.~\hyperlink{supp3}{III}, while in
Sec.~\hyperlink{supp4}{IV} we compare runtimes of the exact and greedy
optimization methods for finding $\mathcal{R}^*$, showing that the
greedy scheme is considerably faster especially for larger
hypergraphs.  Finally, we explore the structural and dynamical
properties of all reduced empirical hypergraphs in
Sec.~\hyperlink{supp7}{VII}. We find that the reduced hypergraph
representations consistently preserve the global, mesoscale, and local
connectivity of the complete empirical hypergraphs---as quantified by
the effective number of connected components, community structure, and
degree ordering, respectively. We also find that these reduced systems
preserve the consensus times in higher-order voter model
dynamics~\cite{kim2025competition}. In general, we observe that such
properties are better preserved as the reducibility $\eta$ increases,
due to improved compressibility of the original hypergraph structure. 

\textit{Conclusion}---Reducing the dimensionality of higher-order
systems allows for more efficient analyses, with simpler
interpretations and visualization.  Here we have developed a
principled, efficient, and interpretable information-theoretic
framework for assessing the structural reducibility of hypergraphs and
removing structural redundancies to construct compressed hypergraph
representations retaining the critical higher-order structure of
complex networked systems. There are number of ways in which this
framework can be extended in future work to directed, weighted,
temporal, or multilayer hypergraphs. This would allow the method to be
applied to representations that capture additional nuances of the
relational structure in a wider variety of systems. Our work sheds new
light on the organizational principles of higher-order networks,
distinguishing the extent to which lower-order information is
redundant in the presence of higher-order information.

\textit{Acknowledgments}---A.K.~acknowledges support from the National
Science Foundation of China~(NSFC) through Young Scientist Fund
Project No.~12405044. F.B.~acknowledges support from the Austrian
Science Fund~(FWF) through projects~10.55776/PAT1052824 and
10.55776/PAT1652425.

\textit{Data availability}---The data that support the findings of
this Letter are openly available~\cite{Landry_XGI_A_Python_2023}. Code
is available as part of the library
Hypergraphx~\cite{lotito2023hypergraphx}.

%
%

\clearpage

\onecolumngrid

\begin{center}
  \textbf{\large Supplemental Material for: \\ \vspace{0.25cm}
Structural reducibility of hypergraphs \vspace{0.25cm}} \\[.2cm]

  Alec Kirkley,$^{1,2,3}$ \, Helcio Felippe,$^4$ and Federico Battiston$^4$ \\ [.1cm]
  {\itshape ${}^1$Institute of Data Science, University of Hong Kong, Hong Kong SAR, China \\
            ${}^2$Department of Urban Planning and Design, University of Hong Kong, Hong Kong SAR, China \\
            ${}^3$Urban Systems Institute, University of Hong Kong, Hong Kong SAR, China \\
            ${}^4$Department of Network and Data Science, Central European University, 1100 Vienna, Austria \\
            }
\end{center}

\setcounter{equation}{0}
\setcounter{figure}{0}
\setcounter{table}{0}
\setcounter{page}{1}
\setcounter{section}{0}
\renewcommand{\theequation}{S\arabic{equation}}
\renewcommand{\thefigure}{S\arabic{figure}}
\renewcommand{\thetable}{S\arabic{table}}
\renewcommand{\thepage}{S\arabic{page}}
\renewcommand{\thesection}{S\arabic{section}}
\renewcommand{\bibnumfmt}[1]{[#1]}
\renewcommand{\citenumfont}[1]{#1}
 
\hypertarget{supp1}{
\section{I. Multiscale Reducibility}\label{section:multiscale}
}

One can extend our formalism to compute a \emph{multiscale} structural
reducibility measure~$\eta(\bm{b})$ for $G$ that considers the extent
to which the hypergraph can be compressed when we coarse-grain $G$
according to an arbitrary node partition~$\bm{b}$, which assigns each
node~$i$ to a group~$b_i$. Such coarse-grained representations are
useful when studying systems with natural communities or node metadata
by which the nodes can be grouped to understand the mesoscale
structure of the system~\cite{felippe2024network}. 

The multiscale measure requires us to consider a coarse-grained
representation of $G$ under the partition $\bm{b}$, which we denote
$\tilde G(\bm{b})$. $\tilde G(\bm{b})$ is a multiset of size~$\abs{G}$
mapping each tuple $(i,j,\dots,k)\in G$ to a tuple
$(b_i,b_j,\dots,b_k)$ of group labels, sorted to ensure all
permutations are equivalent as before. As $\tilde G(\bm{b})$ may have
repeated elements (hyperedge tuples), it is formally represented as a
multiset. Defining the scale of the hypergraph $G$ to be order
$\text{O}(1)$, the representation $\tilde G(\bm{b})$ captures
coarse-grained structure of $G$ at a scale of order
$\text{O}(B^{-1})$, where $B$ is the number of groups in $\bm{b}$. 

\begin{figure}[b]
\includegraphics[width=0.70\textwidth]{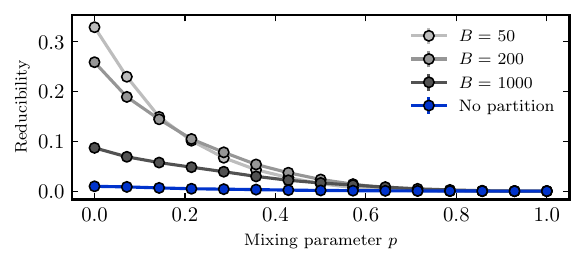}
\caption{
    Multiscale structural reducibility. Multiscale reducibility
    (Eq.~(S6)) versus mixing parameter $p$
    determining the expected fraction of nodes in each hyperedge that
    belong to the majority community within the hyperedge. As $p$
    increases the generated hypergraphs become less clustered with
    respect to the input partition $\bm{b}$. As the number of
    communities $B$ in $\bm{b}$ increases, the hypergraph becomes less
    reducible with respect to $\bm{b}$, approaching the standard
    reducibility (Eq.~(\ref{eq:reducibility}), blue).
}
\label{fig:fig3_meso}
\end{figure}

To transmit a layer $\tilde G^{(r)}(\bm{b})$ of $\tilde G(\bm{b})$, we
must consider the slightly different problem of choosing a multiset of
$E^{(r)}$ tuples from among all $\multiset{B}{r}$ possible unique
hyperedge tuples of size $r$, where $\multiset{n}{k}={n+k-1\choose k}$
is the multiset coefficient counting the number of unique multisets of
size $k$ that can be constructed from $n$ unique elements. This layer
transmission therefore requires
\begin{align}
\log \multiset{\multiset{B}{r}}{E^{(r)}}    
\end{align}
bits of information. We can also adapt our notion of layer overlap to
the multisets $\tilde G^{(k)}(\bm{b})$ and $\tilde G^{(\ell)}(\bm{b})$
by defining the overlap
\begin{align}
E^{(k\cap \ell)}(\bm{b}) = \abs{\tilde G^{(k\to \text{min}(k,\ell))}(\bm{b}) \cap_m \tilde G^{(\ell\to \text{min}(k,\ell))}(\bm{b})},   
\end{align}
where $\cap_m$ is the multiset intersection defined such that the
multiplicity of a tuple $t$ in $A \cap_m B$ is equal to the minimum of
its multiplicity in $A$ and its multiplicity in $B$. We also have used
the notation $\tilde G^{(k\to \ell)}(\bm{b})$ to denote the
coarse-grained tuples in the projected layer $G^{(k\to \ell)}$.

Given knowledge of a coarse-grained representative layer $\tilde
G^{(r(\ell))}(\bm{b})$, we can then transmit $\tilde
G^{(\ell)}(\bm{b})$ using a similar procedure as before, accounting
for the fact that we are working with multisets rather than simple
sets.  $E^{(r(\ell)\cap \ell)}(\bm{b})$ out of the $E^{(r(\ell)\to
\ell)}$ tuples in $G^{(r(\ell)\to \ell)}(\bm{b})$ are shared with
$G^{(\ell)}(\bm{b})$, and there are $\multiset{B}{\ell}$ possible
hyperedges from which the remaining $E^{(\ell)}-E^{(r(\ell)\cap
\ell)}(\bm{b})$ tuples in $G^{(\ell)}(\bm{b})$ must be chosen. We
then have that the information content required for transmitting
the coarse-grained hypergraph $\tilde G(\bm{b})$ using a set of
representative (coarse-grained) layers $\mathcal{R}$ is
\begin{align}\label{eq:dl-multiscale}
H_G(\mathcal{R}\vert \bm{b}) &= \sum_{r\in \mathcal{R}}\log \multiset{\multiset{B}{r}}{E^{(r)}} \\
+ \sum_{\ell\in \mathcal{L}\setminus \mathcal{R}}&\log {E^{(r(\ell)\to \ell)} \choose E^{(r(\ell)\cap \ell)}(\bm{b})}\multiset{\multiset{B}{\ell}}{ E^{(\ell)}-E^{(r(\ell)\cap \ell)}(\bm{b})}.\nonumber
\end{align}
Equation~(\ref{eq:dl-multiscale}) can be minimized over representative
layer subsets $\mathcal{R}$ using the same method as before (with
appropriately modified $M$ and $Q$) to find the optimum
$\mathcal{R}^\ast$.

Defining the quantities
\begin{align}
H_0(\bm{b}) &= \sum_{\ell\in \mathcal{L}}\log \multiset{\multiset{B}{\ell}}{E^{(\ell)}},\\
H_{\ell_{\text{max}}}(\bm{b}) &= \log \multiset{\multiset{B}{\ell_{\text{max}}}}{E^{(\ell_{\text{max}})}},
\end{align}
we can see that $H_{\ell_{\text{max}}}(\bm{b})\leq
H_G(\mathcal{R}^\ast\vert \bm{b})\leq H_0(\bm{b})$ using analogous
arguments to before. A multiscale structural reducibility measure
$\eta(\bm{b})$ for hypergraphs whose nodes are partitioned according
to an arbitrary labelling~$\bm{b}$ is then given by
\begin{align}\label{eq:multiscale-reducibility}
\eta(\bm{b}) = \frac{H_0(\bm{b})-H_G(\mathcal{R}^\ast\vert \bm{b})}{H_0(\bm{b})-H_{\ell_{\text{max}}}(\bm{b})}.    
\end{align}
As before, $\eta(\bm{b})\in [0,1]$ with $\eta(\bm{b})=1$ indicating
maximal hypergraph compressibility under the coarse-graining $\bm{b}$,
and $\eta(\bm{b})=0$ indicating minimal hypergraph compressibility
under the coarse-graining $\bm{b}$.  The multiscale reducibility
reduces to the standard reducibility, adapted for multigraphs, when
each node is in a group by itself. 

In this multiscale framework the choice of node partition $\bm{b}$
used to coarse-grain the network is crucial. Indeed, as shown in
Fig.~\ref{fig:fig3_meso}, as the correlation between the structure of
the hypergraph and the node partition $\bm{b}$ becomes weaker---in
this case, when $p$ is higher so that edges frequently form among
nodes between different communities---the reducibility of the system
decreases. This is because the overlaps that provide structural
redundancy are viewed only from the coarse-grained representation of
the network, and if this coarse-grained representation does not have
any particular structural regularity then little compression is
possible and the reducibility is low. In other words, the more closely
the network structure corresponds to the node labels---i.e. the more
``community-like'' the partition $\bm{b}$ is---the more reducibility
we will see since there will be a greater redundancy of edges with few
node labels. 

Looking at Fig.~\ref{fig:fig3_meso}, we have that in the regime
$p\approx 0$, the node partition $\bm{b}$ corresponds closely with the
edges in the hypergraph---indeed, one could consider the synthetic
model used as a simple generative model for hypergraph community
structure, with the mixing parameter $p$ determining the strength of
community structure. We therefore find high redundancy (e.g.
reducibility) under the partition $\bm{b}$ at low $p$, since edges
tend to form among members of the same community to create lots of
redundant hyperedges under the coarse-grained node labelling. On the
other hand, for $p\to 1$, edges increasingly become composed of
members of different communities, and therefore there is little
redundancy in these hyperedges available to exploit for compression
when $B$ is much larger than the hyperedge size. As $B$ increases,
there become fewer duplicate hyperedges in each layer at low $p$ due
to it being highly probable that a different community is in the
majority at each iteration. We therefore see that the maximum
reducibility for $p=0$ decreases as $B$ increases.

By noting this importance of the coarse-graining, the multiscale
reducibility can then be used to determine the extent to which any
given coarse-graining $\bm{b}$ helps to highlight the structural
redundancies in the hypergraph. One could then in principle examine
multiple partitions $\bm{b}^{(1)}$, $\bm{b}^{(2)}$, etc, each
corresponding to a different set of node metadata, and determine which
set of metadata is most effective in highlighting structural
redundancies in the hypergraph by seeing which partition $\bm{b}$
maximizes the reducibility. One could also in principle search over
partitions $\bm{b}$ that do not correspond to observed node metadata,
to find the coarse-graining under which the hypergraph is most
reducible.

\begin{figure*}[h!]
\centering
\includegraphics[width=0.5\textwidth]{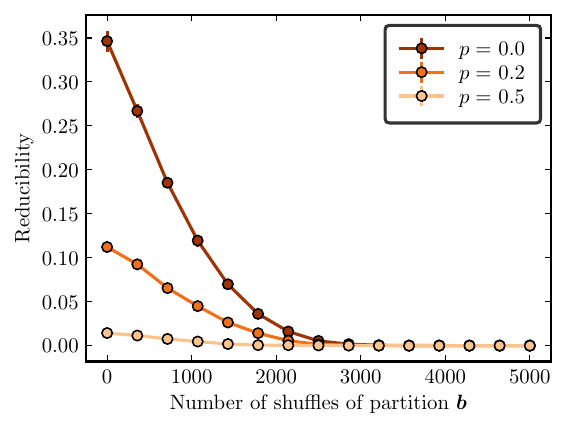}
\caption{
    Impact of node partition on multiscale reducibility. The
    multiscale reducibility of synthetic hypergraphs with community
    structure generated using partition $\bm{b}$ is plotted against
    the amount of noise (number of pairwise shuffles) applied to
    $\bm{b}$ prior to computing the multiscale reducibility. As the
    partition we use for computing the multiscale reducibility becomes
    less correlated with the underlying community structure of the
    graph, we see reducibility drop.  Experiments are repeated over
    ten trials and error bars represent two standard errors in the
    mean.
}
\label{fig:shuffling}
\end{figure*}

In Fig.~\ref{fig:shuffling} we run experiments to systematically
examine the impact that the choice of node partition $\bm{b}$ has on
the reducibility. Specifically, we analyze how the reducibility
changes as the partition we use for the multiscale measure becomes
less and less correlated with the underlying community structure of
the hypergraph $G$ being analyzed.

Using the same synthetic hypergraphs as before, we choose a few fixed
values for the noise $p$, each of which corresponds to different
initial levels of reducibility according to Fig.~\ref{fig:fig3_meso}.
We then add noise to the underlying partition $\bm{b}$ that was used
to generate the network and compute the multiscale reducibility
$\eta(\bm{b}')$ using this shuffled partition $\bm{b}'$. The results
are shown in Fig.~\ref{fig:shuffling}. We can see that, as expected,
the reducibility drops as the partition $\bm{b}'$ we use to compute
the reducibility becomes less and less correlated with the underlying
community structure of the graph (generated by $\bm{b}$).

\hypertarget{supp2}{
\section{II. Reducibility of synthetic hypergraphs with tunable nestedness}\label{appendix:runtime}
}

In this section, we extend the analysis presented in
Fig.~\ref{fig:fig2_structural-functional} of the main text to more
complex models of synthetic hypergraphs with tunable levels of
nestedness and similarity across layers of interactions. In
particular, we consider the following models of $N=100$ node hypergraphs:
 
\begin{itemize}
    \item Model S1: we generate, independently at random, interactions
    of order 3, 5, and 7.  Interactions of order~2 are generated by
    considering tuples of nodes which are subsets of the tuples
    encoding interactions of order~3, while interactions of order~4
    are generated from subsets of tuples encoding interactions of
    order~5, and interactions of order~6 are generated from subsets of
    the tuples encoding interactions of order~7.  Interactions of
    orders 4, 5, 6, and 7 are rewired with probability $\epsilon$,
    while layers~2 and 3 are kept fixed. 

    \item Model S2: Same as model S1, but layers 6 and 7 are rewired
    while the others are kept fixed.
    
    \item Model S3: we generate, independently at random, interactions
    of order 5, 6, and 7.  Interactions of order~2 are generated by
    considering tuples of nodes which are subsets of the tuples
    encoding interactions of order~5, while interactions of order~3
    are generated from subsets of tuples encoding interactions of
    order~6, and interactions of order 6 are generated from subsets of
    the tuples encoding interactions of order~7.
    All layers are rewired with probability~$\epsilon$.

    \item Model S4: same as model S3, but orders 2, 3, 5, and 6 are
    rewired with probability $\epsilon$ while keeping layers~4 and 7 fixed.

    \item Model S5: same as model S3, but rewiring only layers~2 and 5
    with probability~$\epsilon$ while keeping all the other layers fixed.

    \item Model S6: we generate, independently at random, interactions
    of order 4, 6, and 7.  Interactions of order~2 are generated by
    considering tuples of nodes which are subsets of the tuples
    encoding interactions of order~6, while interactions of order~3
    are generated from subsets of tuples encoding interactions of
    order~4, and interactions of order~5 are generated from subsets of
    the tuples encoding interactions of order~7. All layers are 
    rewired with probability $\epsilon$.

\end{itemize}

\begin{figure*}[h!]
\centering
\includegraphics[width=1.0\textwidth]{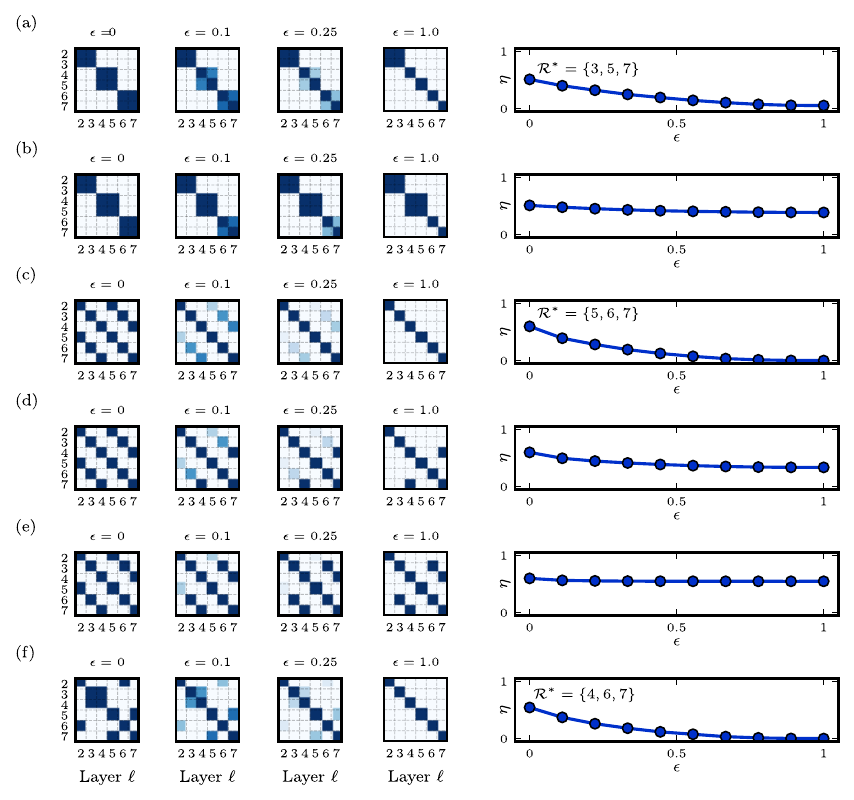}
\caption{
    Reducibility $\eta$ against noise parameter $\epsilon$ for
    synthetic hypergraphs with tunable nestedness.  Heatmaps
    on the left illustrate the pairwise layer similarity at four values of
    $\epsilon$, while the plots on the right-side show the structural
    reducibility $\eta$ against all $\epsilon\in[0,1]$.
    (a)~Model S1 where layers $\ell=2$, 4, and 6 are generated from
    $\ell=3$, 5, and 7, respectively, and all but layers 2 and 3 are
    rewired. The optimal representatives
    $\mathcal{R}^*=\{3,5,7\}$ are precisely obtained at $\epsilon=0$,
    with the reducibility $\eta$ decreasing with $\epsilon$ but never
    reaching zero, as the redundancy between nested layers 2 and 3 is
    kept intact.
    (b)~Model S2, which is equivalent to S1, but layers $\ell=2$, 3,
    4, and 5 are kept fixed (i.e. only $\ell=6$ and 7 are rewired with
    probability~$\epsilon$).  Reducibility drops only slightly from its initial
    value because the layer redundancies are preserved through most
    layers being kept fixed.
    (c)~Model S3, where layers $\ell=5$, 6, and 7 generate
    $\ell=2$, 3, and 4, respectively. All layers are
    rewired with probability $\epsilon$.  As noise levels increase, 
    the reducibility $\eta$ decreases
    and eventually reaches zero because all layers are fully rewired
    at $\epsilon=1$.
    (d)~Model S4, which is equivalent to S3, but layers 4 and 7 are
    kept fixed while all the others are rewired.  Reducibility is
    decreased but remains strictly positive due to left-over layer
    redundancies.
    (e)~Model~S5, which is equivalent to S3, but layers $\ell=3$, 4,
    6, and 7 are kept fixed. 
    Reducibility is decreased but remains close to its
    original value because of the remaining layer redundancies. 
    (f)~Model S6, where layers~$\ell=4$, 6, and 7 are nested within
    layers 3, 2, and 5, respectively. Reducibility $\eta$ vanishes 
    completely because all layers are fully rewired at $\epsilon=1$. 
}
\label{fig:figS1_models}
\end{figure*}

In all of the above models, the density of hyperedges is kept
meaningful across all layers of interactions, in the sense that the
size of layer $\ell$ is set at $E^{(\ell)}=E^{(\ell_{\rm
max})}\binom{\ell_{\rm max}}{\ell}$ with a choice of $E^{(\ell_{\rm
max})}\geq 100$.  In Fig.~\ref{fig:figS1_models} we show the
structural reducibility $\eta$ as a function of the noise parameter
$\epsilon$, for all synthetic scenarios listed. We also show heatmaps
illustrating the hypergraph ``nestedness'' via a measure of similarity
between pairs of orders of interactions, which is encoded in a matrix
with entry values~$\text{I}_{\ell\ell'}$ equal to
\begin{align}\label{eq:Igraph}
\text{I}_{\ell\ell'} = \text{H}_b(p_\ell)+\text{H}_b(p_{\ell'})-\text{H}_s(\bm{P}_{\ell\ell'}),
\end{align}
where $\ell'>\ell$, and
\begin{align}
\bm{P}_{\ell\ell'} = \{p_{\ell\ell'},p_{\ell}-p_{\ell\ell'},p_{\ell'}-p_{\ell\ell'},1-p_{\ell}-p_{\ell'}+p_{\ell\ell'}\}    
\end{align}
is a vector totaling the overlaps among the layers, with
$p_{\ell}=\abs{G^{(\ell)}}/{N\choose \ell}$,
$p_{\ell'}=\abs{G^{(\ell'\to\ell)}}/{N\choose \ell}$, and
\mbox{$p_{\ell\ell'}=\abs{G^{(\ell'\to \ell)}\cap
G^{(\ell)}}/{N\choose \ell}$}.  $\text{H}_b(p)$ and
$\text{H}_s(\bm{x})$ are the binary and Shannon entropies,
respectively. This method is a generalization to $\ell\geq 2$ of the
network mutual information measure for graph similarity developed
in~\cite{felippe2024network}. In all such cases, dynamical
reducibility~\cite{lucas2024functional} (not shown) is unable to
detect the nuances of the structural data, similarly to the example
illustrated in Fig.~\ref{fig:fig2_structural-functional} in the main
text.

\clearpage

\hypertarget{supp3}{
\section{III. The nested organization of real-world higher-order networks}
}

For all real-world systems reported in Table~\ref{tab:datasets}, here
we further investigate their nested organization by displaying their
layer-similarity matrices.  For each pair of layers, the entry of the
similarity matrix is computed via Eq.~(S7).  For
simplicity, we only visualize the first 25 layers of interactions.
The non-zero off-diagonal entries of such matrices reveal a highly
nested structure for many real-world systems, explaining their
structural reducibility.

\begin{figure*}[h!]
\centering
\includegraphics[width=\textwidth]{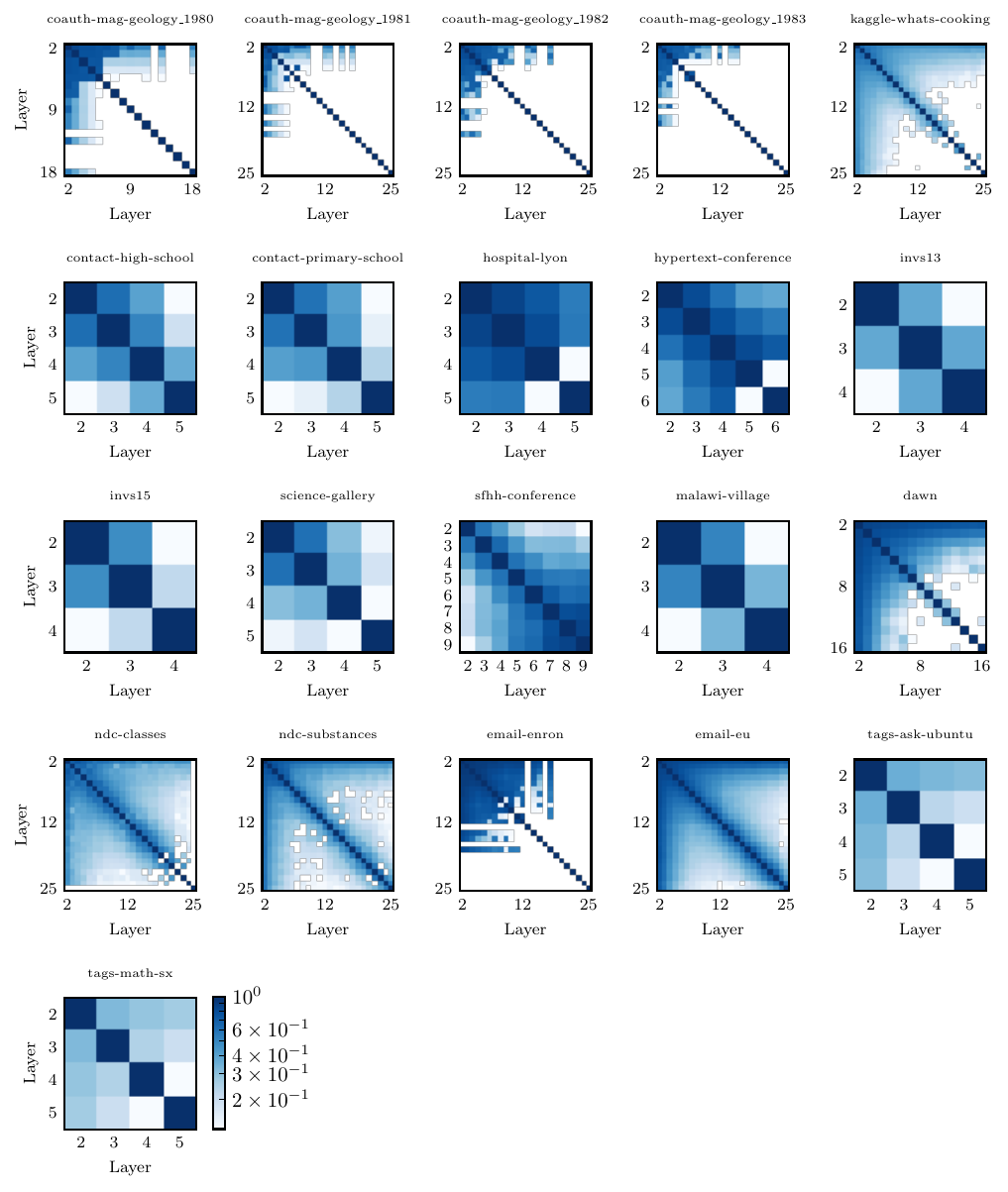}
\caption{
    Nested architecture of empirical hypergraphs.  Color intensity is
    displayed on a log scale.
}
\label{fig:figS3_empirical-heatmaps}
\end{figure*}

\clearpage

\hypertarget{supp4}{
\section{IV. Accuracy and efficiency of the greedy optimization scheme}
}

In this section, we comment on the accuracy and runtime of the
structural reducibility method by performing experiments involving
synthetic and empirical hypergraphs. 

\begin{figure*}[h!]
\centering
\includegraphics[width=0.50\textwidth]{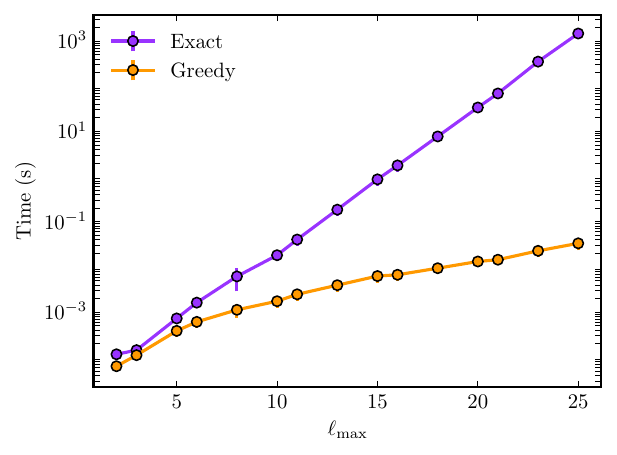}
\caption{
    Runtimes for computing $\mathcal{R}^\ast$ and $\eta$ in synthetic
    fully nested hypergraphs, using both the exact and greedy
    optimization schemes.
}
\label{fig:figS2_exact-greedy}
\end{figure*}

First, we measure the runtime of the exact and greedy optimization
schemes---the time it takes to search and find the optimal
representative set $\mathcal{R}^*$ after computing $M$ and $Q$.  In
Fig.~\ref{fig:figS2_exact-greedy} we display the runtime in seconds
against the maximum number of layers $\ell_{\rm max}$ of fully nested
hypergraphs, as those illustrated in Fig.1 in the main text.  The
simulations are averaged across ten trials and error bars indicate 2
standard errors. The greedy method is always faster than (and quickly
diverges from) the exact scheme, with its advantage clearly shown in
the regime of high-dimensional hypergraphs ($\ell_{\rm max}\geq 15$).
The exact and greedy optimization schemes output the exact same values
for the representative set $\mathcal{R}^*$ and the reducibility $\eta$
in all examples tested.

\begin{figure*}[h!]
\centering
\includegraphics[width=0.50\textwidth]{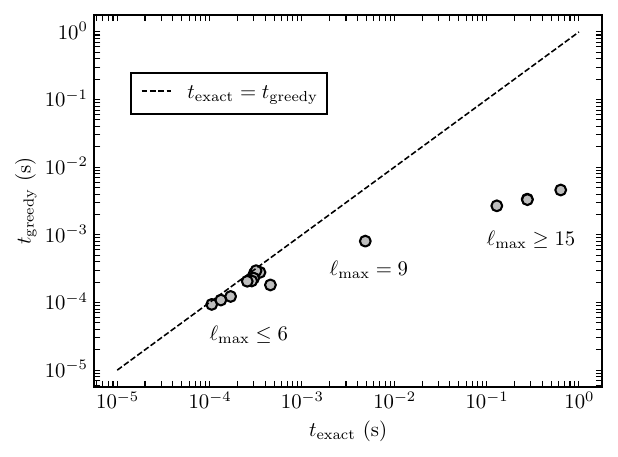}
\caption{
    Runtimes for computing $\mathcal{R}^\ast$ and $\eta$ in empirical
    hypergraphs, using both the exact and greedy optimization schemes.
}
\label{fig:figS2_empirical-exact-greedy}
\end{figure*}

We next analyzed the runtime of both optimization schemes by selecting
a subset of the empirical systems listed in Table~\ref{tab:datasets}
of the main text. In particular, we considered the datasets for which
running the exact optimization scheme was computationally tractable
(datasets with $\ell_{\text{max}}\leq 30$), and compared the runtime
of exact method with the greedy method.
Figure~\ref{fig:figS2_empirical-exact-greedy} shows the runtime
comparison in seconds. Our results align closely with what we observed
for the synthetic simulations, with the exact method's runtime
diverging from the greedy method's runtime in the expected
exponential manner.  This divergence becomes particularly apparent
for $\ell_{\text{max}}\geq 15$, where the greedy runtimes are on
the order of milliseconds while the exact runtimes are on the
order of seconds. Both methods again gave exactly the same
results in all instances studied, further confirming the
reliability of the greedy approach, which allows to extend our
approach to otherwise untractable datasets. 

\hypertarget{supp5}{
\section{V. Individual layer reducibility}
}

Using the information cost in Eq.~(\ref{eq:HGR}), one can also develop
a reducibility measure that assesses each \emph{individual layer}'s
reducibility. This gives us an idea about how redundant any given
layer $\ell$ is in the context of the whole hypergraph.

The information cost to transmit layer $\ell$ can be obtained from the
total information cost in Eq.~(\ref{eq:HGR}) as
\begin{align}\label{eq:costofl}
\log {E^{(r(\ell)\to \ell)} \choose E^{(r(\ell) \cap \ell)}}{{N\choose \ell} - E^{(r(\ell)\to \ell)} \choose E^{(\ell)}-E^{(r(\ell) \cap \ell)}}    
\end{align}
bits, where $r(l)>\ell$ is the representative of layer $\ell$. This
means that the minimum cost of layer $\ell$ over any possible
configuration of representative layers $\mathcal{R}$ (excluding
$\ell$) would be Eq.~(S9) evaluated at
\begin{align}
r^{\ast}(l) = \argmin_{r> \ell\in \mathcal{L}} \left\{\log {E^{(r\to \ell)} \choose E^{(r \cap \ell)}}{{N\choose \ell} - E^{(r\to \ell)} \choose E^{(\ell)}-E^{(r \cap \ell)}} \right\},    
\end{align}
where this time $r$ is allowed to check the whole set of layers in
$\mathcal{L}$.

Meanwhile, the layer $\ell$ would cost us
\begin{align}
\log {{N\choose \ell} \choose E^{(\ell)}}     
\end{align}
bits to transmit without a representative reduced hypergraph to aid in
compression. Therefore, the maximum fractional amount layer $\ell$
could possibly be compressed under any representative hypergraph is
\begin{align}
\eta_{\ell} = 1-\frac{\log {E^{(r^{\ast}(\ell)\to \ell)} \choose E^{(r^{\ast}(\ell) \cap \ell)}}{{N\choose \ell} - E^{(r^{\ast}(\ell)\to \ell)} \choose E^{(\ell)}-E^{(r^{\ast}(\ell) \cap \ell)}}}{\log {{N\choose \ell} \choose E^{(\ell)}}}.     
\end{align}
This measure satisfies $\eta_{\ell}\in [0,1]$, with a maximum
reducibility of $\eta_{\ell}=1$ if and only if layer $\ell$ is a
projection of some other layer $r^{\ast}(\ell)$ in the
hypergraph---i.e., is fully redundant---and $\eta_{\ell} \approx 0$ if
layer $\ell$ has very little overlap with any other layer. We also
require the convention that for the layer of maximum order,
$\ell=\ell_{\text{max}}$, we have $\eta_{\ell}=0$, since these
hyperedges cannot be transmitted from hyperedges of lower order (hence
must be a representative layer). 

\begin{figure}[htb!]
\centering
\includegraphics[width=1.0\textwidth]{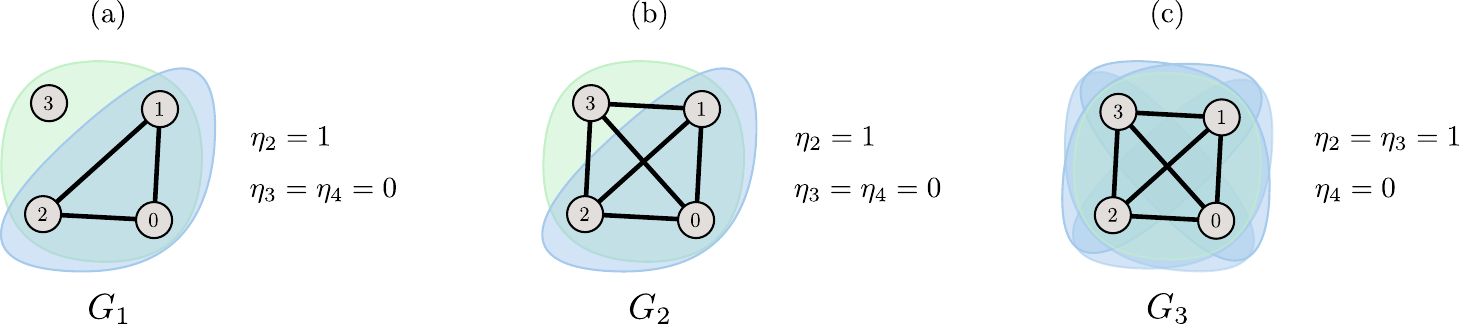}
\caption{
    Small example networks to illustrate layer-wise reducibility.
    Edges are represented as thick lines, 3-body interactions are blue
    rounded triangles, and 4-body interactions are green rounded
    squares.
    (a)~Hypergraph $G_1$, with layer $\ell=2$ fully redundant given
    $\ell=3$, but $\ell=3$ not redundant.
    (b)~Hypergraph $G_2$, with layer $\ell=2$ fully redundant given
    $\ell=4$, but $\ell=3$ again not redundant.
    (c)~Fully nested hypergraph $G_3$, with layer $\ell=2$ fully
    redundant given $\ell=3,4$ and layer $\ell=3$ fully redundant
    given $\ell=4$.     
}
\label{fig:toymodels}
\end{figure}

We can examine this measure on a few small example networks to better
understand its behavior:
\begin{itemize}
    \item $G_1 = \{(0,1,2,3),(0,1,2),(0,1),(0,2),(1,2)\}$. 

    \begin{itemize}
        \item Layer $\ell=2$ is fully redundant as it is a projection $3\to 2$ of $\ell=3$. Therefore we have $\eta_2=1$. 
        \item Layer $\ell=3$ has reducibility $\eta_{3}=0$, as it is less efficient to transmit layer $\ell=3$ from layer $\ell=4$ than it is to transmit layer $\ell=3$ by itself.
        \item Top layer $\ell=4$ has reducibility $\eta_4=0$ by convention, as it cannot be transmitted from a lower layer.
    \end{itemize}
    
    \item $G_2 = \{(0,1,2,3),(0,1,2),(0,1),(0,2),(0,3),(1,2),(1,3),(2,3)\}$.

    \begin{itemize}
        \item Layers $\ell=3$ and $\ell=4$ have the same reducibility values as they are unchanged and cannot be transmitted from layer $\ell=2$.
        \item Layer $\ell=2$ still has reducibility $\eta_2=1$, as it is a projection $4\to 2$ of $\ell=4$.
    \end{itemize}

    \item $G_3 = \{(0,1,2,3),(0,1,2),(0,1,3),(0,2,3),(1,2,3),(0,1),(0,2),(0,3),(1,2),(1,3),(2,3)\}$. $G_3$ is the fully nested hypergraph on $(0,1,2,3)$.

    \begin{itemize}
        \item Layer $\ell=2$ is fully redundant as it is a projection $3\to 2$ of $\ell=3$. Therefore we have $\eta_2=1$. 
        \item Layer $\ell=3$ is fully redundant as it is a projection $4\to 3$ of $\ell=4$. Therefore we have $\eta_3=1$.
        \item Top layer $\ell=4$ has reducibility $\eta_4=0$ again by convention.
    \end{itemize}
    
\end{itemize}
Figure~\ref{fig:toymodels} illustrates the simple examples described
above.

\begin{figure*}[htb!]
\centering
\includegraphics[width=1\textwidth]{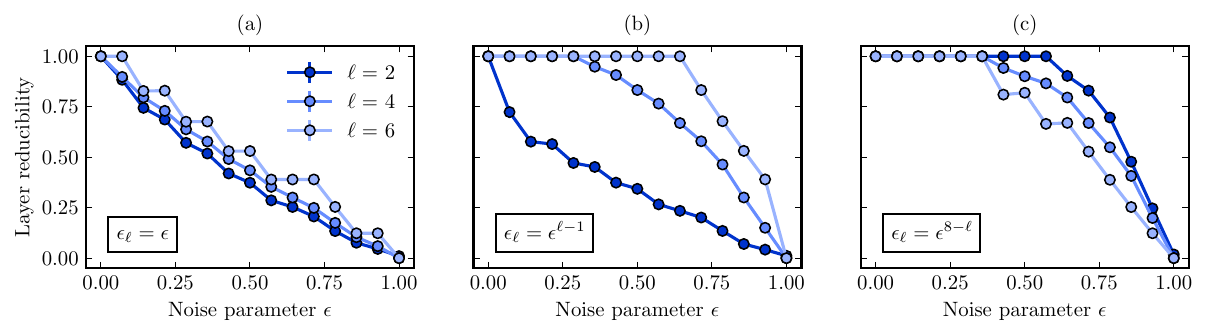}
\caption{
    Layer-wise reducibility of noisy synthetic hypergraphs. Individual
    layer reducibility values for $\ell=2,4,6$ versus noise level
    $\epsilon$ for: (a) Fully nested hypergraphs with constant noise
    schedule $\epsilon_\ell=\epsilon$; (b) Fully nested hypergraphs
    with noise schedule $\epsilon_\ell=\epsilon^{\ell-1}$ more heavily
    perturbing lower order layers; (c) Fully nested hypergraphs with
    noise schedule $\epsilon_\ell=\epsilon^{8-\ell}$ more heavily
    perturbing higher order layers.  Experiments are repeated over ten
    trials and error bars represent two standard errors in the mean.   
}
\label{fig:layerwise}
\end{figure*}

We can also examine how noise heterogeneity among hypergraph layers
can impact the layer-wise reducibility. To test this, we generate
synthetic fully nested hypergraphs via the same mechanism as in
Fig.~\ref{fig:fig1_diagram_eta} in the main text, with
$\ell_{\text{max}}=7$, except this time allow the noise level to vary
across different layers according to a function $\epsilon_l$ that
follows three different noise schedules:

\begin{enumerate}
    \item $\epsilon_\ell=\epsilon$ (Fig.~\ref{fig:layerwise}a). Here, layers will all receive the same amount of noise.
    \item $\epsilon_\ell=\epsilon^{\ell-1}$ (Fig.~\ref{fig:layerwise}b). Here, lower order layers will receive more noise than higher layers for all $\epsilon\in (0,1)$.
    \item $\epsilon_\ell=\epsilon^{8-\ell}$ (Fig.~\ref{fig:layerwise}c). Here, higher order layers will receive more noise than lower layers for all $\epsilon\in (0,1)$.
\end{enumerate}
In Fig.~\ref{fig:layerwise} we show the results of these tests, which
confirm two intuitive expectations: (1) the reducibility of the
individual layers goes from $1$ to $0$ smoothly as we increase the
noise; and (2) as the noise increases across levels, the reducibility
decreases.

\medskip

\begin{figure*}[htb!]
\centering
\includegraphics[width=1\textwidth]{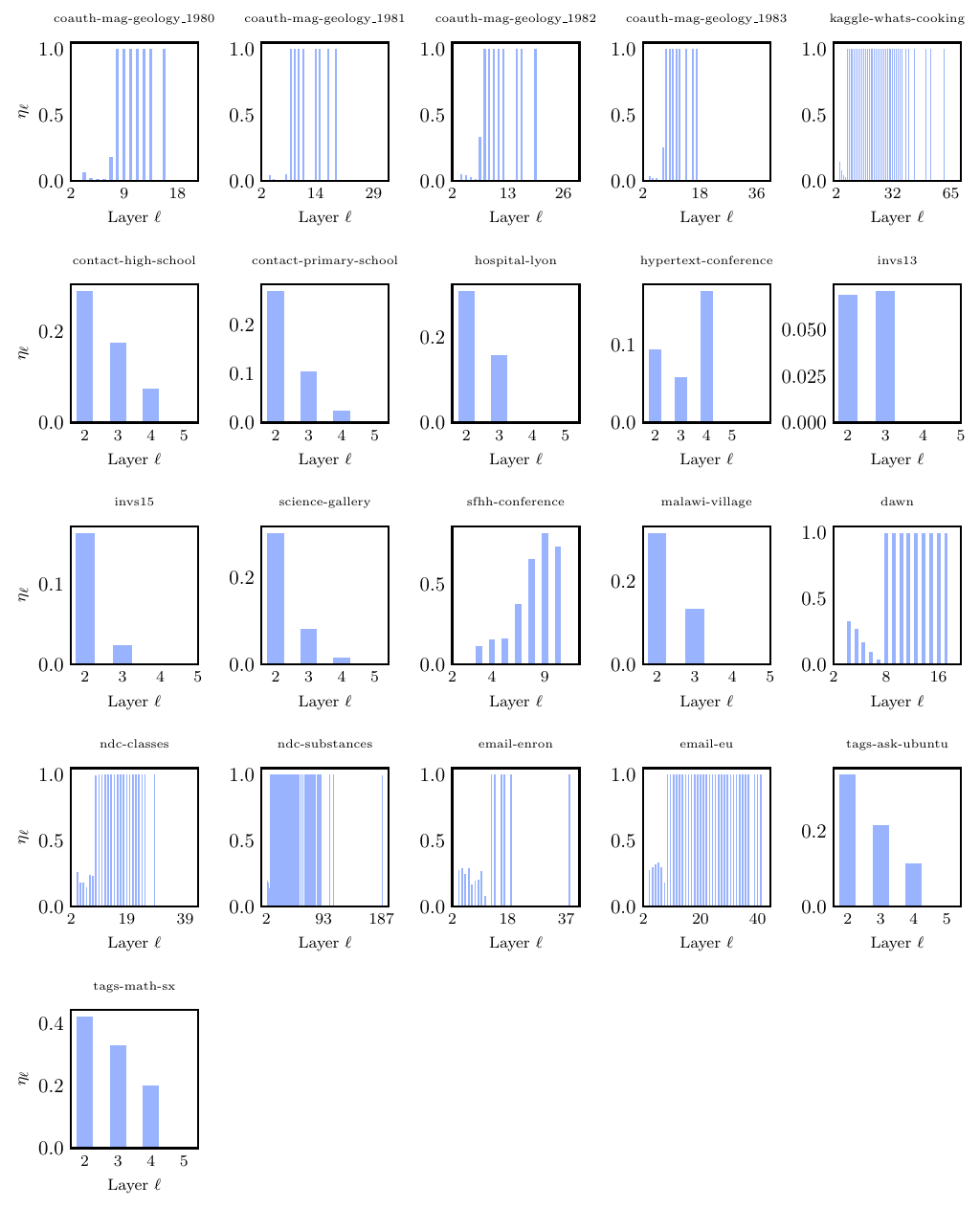}
\caption{
    Layer-wise reducibility of empirical hypergraphs. 
}
\label{fig:layerwise_empirical}
\end{figure*}

Finally, in Figure~\ref{fig:layerwise_empirical} we show the results
of applying the layer-wise reducibility to the empirical hypergraphs
from Table~\ref{tab:datasets}. We can generally see that the
highest-order layers in many hypergraphs are quite reducible
($\eta_\ell \approx 1$), indicating lots of redundancy in the large
hyperedges. However, referencing the reducibility values in Table I in
the main text, we can observe that the overall reducibility $\eta$ is
primarily influenced by the lower-order layers since these contain
most of the hyperedges. For example, the coauth-mag-geology datasets,
which had the lowest total reducibility values, have high reducibility
for their highest-order layers, but reducibility near zero for their
lowest-order layers.

\clearpage

\hypertarget{supp6}{
\section{VI. Reducibility through representative hyperedges}
}

The method presented in this paper aims to compress hypergraphs by
identifying a representative subset of layers $\mathcal{R}\subseteq
\mathcal{L}$ that parsimoniously captures the structural regularities
in the hypergraph. Each of the representative layers serves as a
summary for the lower order layers for which it is a representative.

In principle, this concept could be extended in a more local fashion,
so that instead of inferring full representative layers we focus on
inferring individual representative hyperedges. In this case, a
hyperedge of higher order for which many lower-order hyperedges are a
subset could be a good candidate as a representative hyperedge. We
explore this concept here, finding that it introduces substantial
additional complexities that may prohibit efficient computation in
practice.

In this scenario, instead of searching over subsets
$\mathcal{R}\subseteq \mathcal{L}$ of layers, one would search over
subsets $R\subseteq G$ of the hyperedges in $G$ to minimize an
information cost for transmitting the hypergraph $G$ using $R$ as a
reduced hypergraph. A reasonable objective would have two terms
analogous to those in Eq.~(\ref{eq:HGR}): (1) Transmit the
representative hyperedges $R$; (2) Transmit the remaining hyperedges
$e$ by assigning each to a representative $r(e)\in R$, and then
transmitting the hyperedge $e$ based on its overlap with $r(e)$. This
is identical to the objective of this paper (Eq.~(\ref{eq:HGR})) if
each edge is assigned its own unique layer index $\ell$.

While such a formulation is conceptually appealing, it faces
computational challenges that exceed those encountered by the method
of this paper. Looking at Table~\ref{tab:datasets}, typical empirical
hypergraphs have only $L\lesssim 30$ layers, meaning an exhaustive
search over the $2^{L}$ subsets of these layers is feasible. For the
few examples with a larger number of layers (see examples with a
$\dag$ in the table), it is fortunate that a simple greedy
optimization procedure can quickly find the global optimum of our
reducibility measure in practice with a runtime of $O(L^3)$: For each
of $L-1$ rounds, choosing the best remaining candidate $\ell^\ast \in
\mathcal{L}\setminus \mathcal{R}$ to add to $\mathcal{R}$, we have to
search through $L-R$ layers $\ell\in \mathcal{L}\setminus\mathcal{R}$
and recompute the objective, which in turn requires checking each
other $\ell'\neq \ell\in \mathcal{L}\setminus \mathcal{R}$ to see
whether $\ell'$ will change to $\ell$ as a representative. Summing
over all $L-1$ rounds of the algorithm gives a total runtime
complexity of
\begin{align}
\sum_{R=0}^{L-2}(L-R)(L-R-1) = \sum_{k=0}^{L-2}(k+2)(k+1) \sim O(L^3).   
\end{align}
This runtime complexity is more than sufficient for applying our
proposed method to large empirical hypergraphs, since $L<1000$ in all
cases we could find.

However, if such a greedy procedure were to be used for a \emph{local}
objective such as the one described above, the runtime complexity
would become \emph{substantially} worse, because we would now have
$L=\abs{G}$. We therefore would have a complexity of $O(\abs{G}^3)$
for this ``fast'' greedy search over $R\subseteq G$, which prohibits
the method from being applied to most of the systems we have
studied here. Due to its substantially increased complexity, we
leave further exploration of this idea to future work. 

\clearpage

\hypertarget{supp7}{
\section{VII. Structural and dynamical properties of reduced hypergraphs}
}
In this supplement, we explore the potential for the reduced
hypergraph representation of $G$, given by
$G_{\text{red}}=\{G^{(r)}\}_{r\in \mathcal{R}^\ast}$, to preserve
structural and dynamical properties of the original hypergraph $G$.
This allows us to potentially use $G_{\text{red}}$ as a sparsified
representation of $G$ to simplify and speed up various computations.

We first examine the structural properties of the original hypergraphs
relative to their reduced counterparts. In Table~\ref{tab:redstats} we
report the number of nodes $N$, number of edges $E=\abs{G}$ in the
original hypergraph, and number of edges
$E_{\text{red}}=\abs{G_{\text{red}}}$ in the reduced version of the
hypergraph, for each empirical dataset studied in the main text. We
can see a substantial reduction in edges for most datasets, consistent
with the reducibility results that indicate only a small subset of
layers $\mathcal{R}^\ast$ are MDL-optimal for the reduced
representation. Given this substantial reduction in the number of
hyperedges, the reduced representations naturally cannot preserve
certain structural and dynamical properties that depend on edge
density or degree distributions---this is also true of any other
network sparsification procedure that removes a sizable number of
edges. However, it is worthwhile to examine structural properties that
are not heavily dependent on the number of edges retained, as
maintenance of these features can allow for easier usage of the
reduced hypergraphs in downstream applications. Here we compute three
key structural properties that provide summaries of the extent to
which the global, mesoscale, and local topologies of the hypergraphs
are disrupted after reduction: the effective number of connected
components in the reduced hypergraph (global), the partitions inferred
by a community detection algorithm run on both the original and
reduced hypergraphs (mesoscale), and the Pearson correlation between
the degree distributions of the original and reduced hypergraphs
(local).
 
\begin{table}[htb!]
    \centering
    \scriptsize
    \rowcolors{2}{white}{gray!10}
    \begin{tabular}{lcccccccccc} 
    \toprule 
    \textbf{Dataset} & {$N$} & {$E$} & $E_{\text{red}}$ & $C_{\text{eff}}(G_{\text{red}})$ & $\text{AMI}(\bm{b},\bm{b}_{\text{red}})$ & $\rho(\bm{k}_{G},\bm{k}_{G_\text{red}})$ & $\log_{10}\tau^{(1)}$ & $\log_{10}\tau^{(1)}_{\text{red}}$ & $\log_{10}\tau^{(2)}$ & $\log_{10}\tau^{(2)}_{\text{red}}$ \\
    \midrule 
    \fontsize{6}{7}\selectfont{coauth-mag-geology\_1980} & 1674 &  903    &  354  & 80.58 &  0.48(0.005) & 0.49 & $\infty$ & $\infty$ & $\infty$ & $\infty$\\ 
    \fontsize{6}{7}\selectfont{coauth-mag-geology\_1981} & 1075 &  547    &  238  & 46.41 &  0.55(0.007) & 0.32 & $\infty$ & $\infty$ & $\infty$ & $\infty$\\
    \fontsize{6}{7}\selectfont{coauth-mag-geology\_1982} & 1878 &  987    &  357  & 92.53 &  0.47(0.005) & 0.44 & $\infty$ & $\infty$ & $\infty$ & $\infty$\\ 
    \fontsize{6}{7}\selectfont{coauth-mag-geology\_1983} & 1734 &  883    &  352  & 77.82 &  0.50(0.005) & 0.20 & $\infty$ & $\infty$ & $\infty$ & $\infty$\\
    \fontsize{6}{7}\selectfont{kaggle-whats-cooking}     & 6714 & 39224   & 13305 & 1.0   &  0.40(0.008) & 0.96 & $\infty$ & $\infty$ & 4.97 & 4.81\\ 
    \fontsize{6}{7}\selectfont{contact-high-school}      & 327  &  7818   & 2098  & 1.0   &  0.71(0.005) & 0.86 & 4.72     & 4.70 & 4.18 & 4.41\\ 
    \fontsize{6}{7}\selectfont{contact-primary-school}   & 242  & 12704   & 356   & 1.0   &  0.67(0.008) & 0.70 & 4.37     & 4.39 & 3.90 & 4.18\\ 
    \fontsize{6}{7}\selectfont{hospital-lyon}            & 75   &  1824   & 60    & 1.0   &  0.01(0.02)  & 0.82 & 3.32     & 2.66 & 3.09 & 2.32\\ 
    \fontsize{6}{7}\selectfont{hypertext-conference}     & 113  &  2434   & 313   & 1.0   &  0.04(0.005) & 0.82 & 3.66     & 3.63 & 3.52 & 3.32\\ 
    \fontsize{6}{7}\selectfont{invs13}                   & 92   &  787    & 46    & 1.0   &  0.53(0.03)  & 0.66 & 3.59     & 3.42 & 3.49 & 3.60\\
    \fontsize{6}{7}\selectfont{invs15}                   & 217  &  4909   & 767   & 1.0   &  0.40(0.02)  & 0.72 & 4.24     & 4.28 & 4.09 & 3.78\\ 
    \fontsize{6}{7}\selectfont{science-gallery}          & 410  &  3350   & 813   & 1.20  &  0.75(0.007) & 0.83 & 4.93     & 4.98 & 4.71 & 4.99\\ 
    \fontsize{6}{7}\selectfont{sfhh-conference}          & 403  &  10541  & 260   & 1.19  &  0.36(0.009) & 0.55 & 4.74     & 4.91 & 4.23 & 4.81\\ 
    \fontsize{6}{7}\selectfont{malawi-village}           & 84   &  431    & 90    & 3.91  &  0.66(0.03)  & 0.69 & 3.57     & 4.89 & 3.50 & 4.99\\ 
    \fontsize{6}{7}\selectfont{dawn}                     & 2290 & 138742  & 13243 & 1.0   &  0.28(0.01)  & 0.97 & 4.98     & 4.92 & 4.70 & 4.68\\ 
    \fontsize{6}{7}\selectfont{ndc-classes}              & 628  &  796    & 212   & 2.06  &  0.71(0.009) & 0.95 & $\infty$ & $\infty$ & $\infty$ & $\infty$\\ 
    \fontsize{6}{7}\selectfont{ndc-substances}           & 3414 &  6471   & 502   & 1.41  &  0.17(0.004) & 0.55 & $\infty$ & $\infty$ & $\infty$ & $\infty$\\
    \fontsize{6}{7}\selectfont{email-enron}              & 143  &  1459   & 195   & 1.0   &  0.74(0.02)  & 0.82 & 3.97     & 3.95 & 3.46 & 3.65\\ 
    \fontsize{6}{7}\selectfont{email-eu}                 & 986  &  24520  & 3067  & 1.0   &  0.79(0.007) & 0.91 & 4.97     & 4.92 & 4.47 & 4.41\\ 
    \fontsize{6}{7}\selectfont{tags-ask-ubuntu}          & 3021 &  145053 & 25475 & 1.0   &  0.56(0.003) & 0.98 & $\infty$ & $\infty$ & 4.82 & 4.64\\ 
    \fontsize{6}{7}\selectfont{tags-math-sx}             & 1627 &  169259 & 29244 & 1.0   &  0.65(0.005) & 0.98 & 4.96     & 4.96 & 4.44 & 4.44\\
    \bottomrule
    \end{tabular}
    \caption{Structural and dynamical properties of empirical hypergraphs and their reduced representations.
    }
    \label{tab:redstats}
\end{table}
 
The first measure we examine is the effective number of connected
components \cite{riolo2017efficient,morel2024bayesian}, given by
\begin{align}
C_{\text{eff}} = \text{exp}\left(-\sum_{s=1}^{S}\frac{n_s}{N}\ln \frac{n_s}{N}\right),    
\end{align}
where $n_s$ is the number of nodes in connected component $s$, and $S$
is the number of connected components. In the extreme cases of a
single component and $N$ isolated components, we have
$C_{\text{eff}}=1$ and $C_{\text{eff}}=N$ respectively, and when one
large component dominates in size we have $C_{\text{eff}}\approx 1$.
By computing $C_{\text{eff}}$ for the original hypergraphs $G$ and
their reduced versions $G_{\text{red}}$, we can see to what extent the
reduction procedure disrupts the global network connectivity. 

The second measure we examine is the adjusted mutual information
$\text{AMI}(\bm{b},\bm{b}_{\text{red}})$ \cite{vinh2009information}
between the node partitions inferred for the original and reduced
hypergraphs (denoted with $\bm{b}$ and $\bm{b}_{\text{red}}$
respectively) using the higher order variant of the Infomap community
detection method \cite{eriksson2021choosing}. In order to compare the
partitions in a consistent manner, we only consider nodes present in
both the original and reduced hypergraphs, in case some nodes were not
present in the reduced representation. The adjusted mutual information
is computed by subtracting from the observed mutual information
between the inferred partitions the expected value of the mutual
information given these partitions' group sizes. Thus, an AMI value of
$0$ indicates a level of similarity among the two partitions that is
no greater than one would expect by chance, and a value substantially
higher than $0$ indicates a meaningful level of similarity among the
inferred partitions. To determine whether the inferred partitions
$\bm{b},\bm{b}_{\text{red}}$ have a meaningfully higher value than
zero, we compute the standard deviation of the AMI for 1000 random
permutations of the reduced partition and the original partition. An
AMI value more than two or three standard deviations above zero for
the inferred reduced partition indicates that the mesoscale structure
is preserved in a meaningful way in the reduced hypergraph.

The final measure we examine is the Pearson correlation coefficient
$\rho(\bm{k}_{G},\bm{k}_{G_{\text{red}}})$ between the degree
distributions of the two networks, where the degree $k(i)$ of a node
$i$ in hypergraph $G$ is the number of unique nodes that $i$ is
connected to by at least one hyperedge, and $\bm{k}_G$ collects these
degree values for all nodes in $G$. Thus, although one cannot in
general preserve the actual degree values when sparsifying a network,
this measure identifies the extent to which the reduced hypergraphs
preserve the relative node degrees, allowing us to evaluate the
disruption to the local connectivity across the hypergraph after the
reduction.    

Table~\ref{tab:redstats} shows the results of computing these measures
for all empirical hypergraphs studied. Each original hypergraph in the
corpus has a single component, so $C_{\text{eff}}=1$ and we only
report the value for the reduced hypergraph. We can observe that
the large-scale connectivity of the hypergraphs is largely
retained after reduction, with 15 out of 21 reduced hypergraphs
still consisting of a single connected component or having
$C_{\text{eff}} \approx 1$ due to the giant component occupying
nearly all of the network. It is worth noting that the four
coauth-mag-geology datasets have quite high values of
$C_{\text{eff}}$, indicating that their reduced counterparts are
highly fragmented. These coauth-mag-geology datasets have the
lowest reducibility values among all datasets studied, indicating
that even their MDL-optimal reductions do not provide very
effective compression. This is due many separate connected
components within each layer of the original hypergraph---it was
unlikely for an author to have multiple papers with the exact same
number of coauthors in a single year. As a result, even the
optimal subsets of layers do not effectively retain the global
connectivity of these four hypergraphs.

Table~\ref{tab:redstats} also shows the AMI values
$\text{AMI}(\bm{b},\bm{b}_{\text{red}})$ among the original and
reduced hypergraphs, along with the standard deviations in their
corresponding randomized realizations over 1000 trials (in
parentheses). We can see that for nearly all empirical hypergraphs
studied, the AMI values are significantly higher than expected by
chance, with the exception of the hospital-lyon dataset. (We also see
that the AMI value, though significant, is relatively low ($0.04$) for
the hypertext-conference dataset.) Interestingly, we can see that the
community structure is fairly well preserved for the
coauth-mag-geology datasets, with moderate AMI values of $\approx
0.5$. This is because the separate small components that form in the
reduced hypergraphs of these datasets represent groups of nodes that
were not very well connected in the original dataset anyway---thus,
these groups of nodes are found as individual communities in both the
original hypergraph and in the reduced hypergraph.

Finally, we can see from Table~\ref{tab:redstats} that the relative
levels of local connectivity are consistently maintained across
empirical hypergraphs as well, with high correlation coefficients
across the board, with the exception of the coauth-mag-geology
datasets for the same reasons mentioned above. All together, our
results indicate that although the reduced hypergraphs are much
sparser than the original hypergraphs (by construction), they can
consistently retain key global, mesoscale, and local structural
properties, so long as these properties are not highly sensitive to
edge density.

\begin{figure*}[htb!]
\centering
\includegraphics[width=0.7\textwidth]{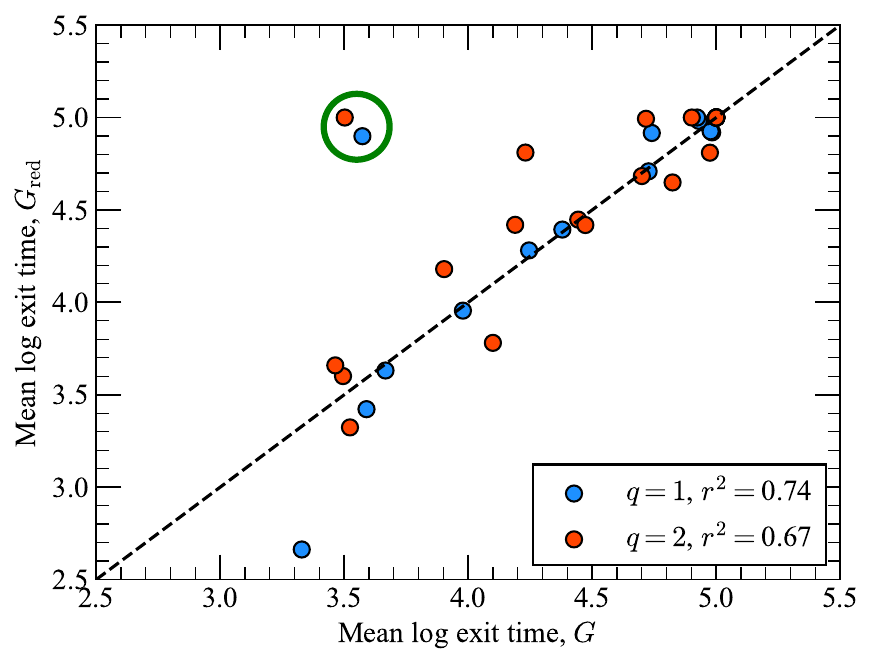}
\caption{
    Mean log exit times for $q=1$ and $q=2$ hypergraph voter model
    dynamics, as listed in Table~\ref{tab:redstats}. The line of
    equality is depicted with the dashed black line, and the
    malawi-village dataset results are highlighted with the green
    circle. Results with $\tau=\infty$ (no consensus over $T=10^{5}$
    simulations) are also plotted, at $(x,y)=(5,5)$.   
}
\label{fig:exits}
\end{figure*}

To examine the extent to which hypergraph \emph{dynamics} can be
preserved after reduction to the representative layers, we study
multiple variants of the voter model on the original and reduced
hypergraphs~\cite{sood2008voter}. The voter model is a widely studied
dynamical model that captures a spectrum of interesting dynamical
behaviors on networks, and as a result has seen many extensions
including to hypergraphs \cite{kim2025competition}. In this model,
each node $i$ starts with a random initial state $\sigma_i(t=0)\in
\{0,1\}$, which represents an initial opinion for node $i$. Then, at
each timestep $t\in \{1,...,T\}$, a random node $i$ is chosen, $i$
picks a random hyperedge of which it is a member, then $i$ chooses
$q\geq 1$ neighbors within this hyperedge (with replacement) at random
and copies the votes of these neighbors if they all agree. A key
observable from this ``group-driven'' voter model process is the
\emph{exit time}, $\tau$, which represents the expected amount of time
it takes for the system to reach a consensus state (either all $0$'s
or all $1$'s)~\cite{kim2025competition}.

We simulate this voter model with 20 independent simulations for each
of the empirical hypergraphs studied in the main text, as well as for
their corresponding reduced hypergraphs. We summarize the results in
Table~\ref{tab:redstats}, where the logarithm of the average exit
times $\tau^{(q)}$ and $\tau_{\text{red}}^{(q)}$ for the original and
reduced hypergraphs are reported for $q=1,2$. $\tau=\infty$ is
assigned to any network which never reaches consensus during any of
the simulations. We can see that the exit time estimates are quite
close for most hypergraph pairs, and of the same order of magnitude
for nearly all hypergraph pairs. This indicates that the reduced
version of each empirical network consistently preserves the voter
consensus dynamics of the original dataset. We can also observe
that for all datasets with $\tau=\infty$ (no consensus after
$T=100,000$ in any of the simulations), the reduced hypergraph
also does not form a consensus in the voter model dynamics. In
Fig.~\ref{fig:exits} we plot the exit times in
Table~\ref{tab:redstats} for easier visualization of the variation
in exit times, along with the corresponding $r^2$ values for the
scatter plots. We also highlight an outlier, the malawi-village
dataset, for which the exit times on the reduced hypergraph are
substantially larger than those for the original hypergraph. This
can be attributed to a breaking up of the original hypergraph to
form multiple connected components ($C_{\text{eff}}=3.91$ for this
dataset), which are unlikely to come to a consensus during the
voting dynamics. Other reduced hypergraphs with $C_{\text{eff}}>1$
either did not reach consensus (as was the case for their original
hypergraph counterpart) or had few enough small components that
consensus could be achieved among all nodes. The lack of consensus
in some of the original hypergraphs reflects their poor global
connectivity and is consistent with the multiple connected
components found in their reduced versions.


\begin{thebibliography}{10}
\expandafter\ifx\csname url\endcsname\relax
  \def\url#1{\texttt{#1}}\fi
\expandafter\ifx\csname urlprefix\endcsname\relax\def\urlprefix{URL }\fi

\bibitem{battiston2020networks}
F.~Battiston, G.~Cencetti, I.~Iacopini, V.~Latora, M.~Lucas, A.~Patania, J.-G. Young, and G.~Petri, 
\href{https://doi.org/10.1016/j.physrep.2020.05.004}
{{Phys. Rep.} \textbf{874}, 1--92 (2020)}

\bibitem{battiston2021physics}
F.~Battiston, E.~Amico, A.~Barrat, G.~Bianconi, G.~Ferraz~de Arruda, B.~Franceschiello, I.~Iacopini, S.~K{\'e}fi, V.~Latora, Y.~Moreno, {et~al.}, 
\href{https://doi.org/10.1038/s41567-021-01371-4}{
{Nat. Phys.} \textbf{17}(10), 1093--1098 (2021)}.

\bibitem{bianconi2021higher}
G.~Bianconi, {Higher-order networks}. Cambridge University Press (2021).

\bibitem{bick2023higher}
C.~Bick, E.~Gross, H.~A. Harrington, and M.~T. Schaub, 
\href{https://doi.org/10.1137/21M1414024}{
{SIAM Rev.} \textbf{65}(3), 686--731 (2023)}.

\bibitem{majhi2022dynamics}
S.~Majhi, M.~Perc, and D.~Ghosh, 
\href{https://doi.org/10.1098/rsif.2022.0043}{
{J. R. Soc. Interface} \textbf{19}(188), 20220043 (2022).}

\bibitem{berge1984hypergraphs}
C.~Berge, {Hypergraphs: Combinatorics of Finite Sets}, volume~45. Elsevier (1984).

\bibitem{benson2018simplicial}
A.~R. Benson, R.~Abebe, M.~T. Schaub, A.~Jadbabaie, and J.~Kleinberg, 
\href{https://doi.org/10.1073/pnas.1800683115}{
{Proc. Natl. Acad. Sci. U.S.A.} \textbf{115}(48), E11221--E11230 (2018)}.

\bibitem{petri2018simplicial}
G.~Petri and A.~Barrat, 
\href{https://doi.org/10.1103/PhysRevLett.121.228301}{
{Phys. Rev. Lett.} \textbf{121}(22), 228301 (2018)}.

\bibitem{contisciani2022inference}
M.~Contisciani, F.~Battiston, and C.~De~Bacco, 
\href{https://doi.org/10.1038/s41467-022-34714-7}{
{Nat. Commun.} \textbf{13}(1), 7229 (2022)}.

\bibitem{di2024percolation}
L.~Di~Gaetano, F.~Battiston, and M.~Starnini, 
Percolation and topological properties of temporal higher-order networks
\href{https://doi.org/10.1103/PhysRevLett.132.037401}{
{Phys. Rev. Lett.} \textbf{132}(3), 037401 (2024)}.

\bibitem{iacopini2019simplicial}
I.~Iacopini, G.~Petri, A.~Barrat, and V.~Latora, 
\href{https://doi.org/10.1038/s41467-019-10431-6}{
{Nat. Commun.} \textbf{10}(1), 2485 (2019)}.

\bibitem{burgio2024triadic}
G.~Burgio, S.~G{\'o}mez, and A.~Arenas, 
\href{https://doi.org/10.1103/PhysRevLett.132.077401}{
{Phys. Rev. Lett.} \textbf{132}(7), 077401 (2024)}.

\bibitem{ferraz2024contagion}
G.~Ferraz~de Arruda, A.~Aleta, and Y.~Moreno, 
\href{https://doi.org/10.48550/arXiv.2402.14938}{
{Nat. Rev. Phys.} \textbf{6}(8), 468--482 (2024)}.

\bibitem{di2024dynamical}
L.~Di~Gaetano, G.~Carugno, F.~Battiston, and F.~Coghi, 
\href{https://doi.org/10.1103/PhysRevLett.133.107401}{
{Phys. Rev. Lett.} \textbf{133}(10), 107401 (2024)}.

\bibitem{skardal2019abrupt}
P.~S. Skardal and A.~Arenas, 
\href{https://doi.org/10.1103/PhysRevLett.122.248301}{
{Phys. Rev. Lett.} \textbf{122}(24), 248301 (2019)}.

\bibitem{millan2020explosive}
A.~P. Mill{\'a}n, J.~J. Torres, and G.~Bianconi, 
\href{https://doi.org/10.1103/PhysRevLett.124.218301}{
{Phys. Rev. Lett.} \textbf{124}(21), 218301 (2020)}.

\bibitem{zhang2023higher}
Y.~Zhang, M.~Lucas, and F.~Battiston, 
\href{https://doi.org/10.1038/s41467-023-37190-9}{
{Nat. Commun.} \textbf{14}(1), 1605 (2023)}.

\bibitem{anwar2024collective}
M.~S. Anwar, G.~K. Sar, M.~Perc, and D.~Ghosh, 
\href{https://doi.org/10.48550/arXiv.2309.03343}{
{Commun. Phys.} \textbf{7}(1), 59 (2024)}.

\bibitem{majhi2024patterns}
S.~Majhi, S.~Ghosh, P.~K. Pal, S.~Pal, T.~K. Pal, D.~Ghosh, J.~Zavr{\v{s}}nik, and M.~Perc, 
\href{https://doi.org/10.1016/j.plrev.2024.12.013}{
{Phys. Life Rev.}  (2024)}.

\bibitem{alvarez2021evolutionary}
U.~Alvarez-Rodriguez, F.~Battiston, G.~F. de~Arruda, Y.~Moreno, M.~Perc, and V.~Latora, 
\href{https://doi.org/10.1038/s41562-020-01024-1}{
{Nat. Hum. Behav.} \textbf{5}(5), 586--595 (2021)}.

\bibitem{civilini2024explosive}
A.~Civilini, O.~Sadekar, F.~Battiston, J.~G{\'o}mez-Garde{\~n}es, and V.~Latora, 
\href{https://doi.org/10.1103/PhysRevLett.132.167401}{
{Phys. Rev. Lett.} \textbf{132}(16), 167401 (2024)}.

\bibitem{kumar2021evolution}
A.~Kumar, S.~Chowdhary, V.~Capraro, and M.~Perc, 
\href{https://doi.org/10.1103/PhysRevE.104.054308}{
{Phys. Rev. E} \textbf{104}(5), 054308 (2021)}.

\bibitem{lotito2022higher}
Q.~F. Lotito, F.~Musciotto, A.~Montresor, and F.~Battiston, 
\href{https://doi.org/10.1038/s42005-022-00858-7}{
{Commun. Phys.} \textbf{5}(1), 79 (2022)}.

\bibitem{larock2023encapsulation}
T.~LaRock and R.~Lambiotte, 
\href{https://doi.org/10.1088/2632-072X/ad0b39}{
{J. Phys. Complex.} \textbf{4}(4), 045007 (2023)}.

\bibitem{landry2024simpliciality}
N.~W. Landry, J.-G. Young, and N.~Eikmeier, 
\href{https://doi.org/10.1140/epjds/s13688-024-00458-1}{
{EPJ Data Sci.} \textbf{13}(1), 17 (2024)}.

\bibitem{gallo2024higher}
L.~Gallo, L.~Lacasa, V.~Latora, and F.~Battiston, 
\href{https://doi.org/10.1038/s41467-024-48578-6}{
{Nat. Commun.} \textbf{15}(1), 4754 (2024)}.

\bibitem{de2015structural}
M.~De~Domenico, V.~Nicosia, A.~Arenas, and V.~Latora, 
\href{https://doi.org/10.1038/ncomms7864}{
{Nat. Commun.} \textbf{6}(1), 6864 (2015)}.

\bibitem{de2016spectral}
M.~De~Domenico and J.~Biamonte, 
\href{https://doi.org/10.1103/PhysRevX.6.041062}{
{Phys. Rev. X} \textbf{6}(4), 041062 (2016)}.

\bibitem{santoro2020algorithmic}
A.~Santoro and V.~Nicosia, 
\href{https://doi.org/10.1103/PhysRevX.10.021069}{
{Phys. Rev. X} \textbf{10}(2), 021069 (2020)}.

\bibitem{kirkley2023compressing}
A.~Kirkley, A.~Rojas, M.~Rosvall, and J.-G. Young, 
\href{https://doi.org/10.1038/s42005-023-01270-5}{
{Commun. Phys.} \textbf{6}(1), 148 (2023)}.

\bibitem{kirkley2022representative}
A.~Kirkley and M.~Newman, 
\href{https://doi.org/10.1038/s42005-022-00816-3}{
{Commun. Phys.} \textbf{5}(1), 40 (2022)}.

\bibitem{lucas2024functional}
M.~Lucas, L.~Gallo, A.~Ghavasieh, F.~Battiston, and M.~De~Domenico, 
\href{https://doi.org/10.48550/arXiv.2404.08547}{
{\textnormal{arXiv:2404.08547}}  (2024)}.

\bibitem{felippe2024network}
H.~Felippe, F.~Battiston, and A.~Kirkley, 
\href{https://doi.org/10.1038/s42005-024-01830-3}{
{Commun. Phys.} \textbf{7}(1), 335 (2024)}.

\bibitem{Landry_XGI_A_Python_2023}
N.~W. Landry, M.~Lucas, I.~Iacopini, G.~Petri, A.~Schwarze, A.~Patania, and L.~Torres, 
\href{https://doi.org/10.21105/joss.05162}{
{J. Open Source Softw.} \textbf{8}(85), 5162 (2023)}.

\bibitem{prlSM}
See Supplemental Material at \href{http://link.aps.org/supplemental/10.1103/xrn7-cz8v}{http://link.aps.org/supplemental/10.1103/xrn7-cz8v},
including references~\cite{kim2025competition,riolo2017efficient,morel2024bayesian,vinh2009information,eriksson2021choosing,sood2008voter}.

\bibitem{kim2025competition}
J.~Kim, D.-S. Lee, B.~Min, M.~A. Porter, M.~San~Miguel, and K.-I. Goh, 
\href{https://doi.org/10.1103/PhysRevE.111.L052301}{
{Phys. Rev. E} \textbf{111}(5), L052301 (2025)}.

\bibitem{riolo2017efficient}
M.~A. Riolo, G.~T. Cantwell, G.~Reinert, and M.~E. Newman, 
\href{https://doi.org/10.1103/PhysRevE.96.032310}{
{Phys. Rev. E} \textbf{96}(3), 032310 (2017)}.

\bibitem{morel2024bayesian}
S.~Morel-Balbi and A.~Kirkley, 
\href{https://doi.org/10.1103/PhysRevResearch.6.033307}{
{Phys. Rev. Res.} \textbf{6}(3), 033307 (2024)}.

\bibitem{vinh2009information}
N.~X. Vinh, J.~Epps, and J.~Bailey, 
\href{https://doi.org/10.1145/1553374.1553511}{
{Proceedings of the 26th Annual International Conference on Machine Learning}, pp. 1073--1080 (2009)}.

\bibitem{eriksson2021choosing}
A.~Eriksson, D.~Edler, A.~Rojas, M.~de~Domenico, and M.~Rosvall, 
\href{https://doi.org/10.1038/s42005-021-00634-z}{
{Commun. Phys.} \textbf{4}(1), 133 (2021)}.

\bibitem{sood2008voter}
V.~Sood, T.~Antal, and S.~Redner, 
\href{https://doi.org/10.1103/PhysRevE.77.041121}{
{Phys. Rev. E} \textbf{77}(4), 041121 (2008)}.

\bibitem{lotito2023hypergraphx}
Q.~F.~Lotito, M.~Contisciani, C.~De~Bacco, L.~Di~Gaetano, L.~Gallo, A.~Montresor, F.~Musciotto, N.~Ruggeri, and F.~Battiston, 
\href{https://doi.org/10.1093/comnet/cnad019}{
{J. Complex Networks} \textbf{11}, cnad019 (2023)}.

\end{thebibliography}
\end{document}